\documentclass[prx,twocolumn,showpacs,psfig,superscriptaddress,longbibliography]{revtex4-2}

\usepackage{amssymb}
\usepackage[ruled,vlined]{algorithm2e}
\usepackage{amsmath}
\usepackage{braket}
\usepackage{booktabs}
\usepackage{chngcntr}
\usepackage{color}
\usepackage{dsfont}
\usepackage{etoolbox}
\usepackage{etex}

\usepackage[T1]{fontenc}
\usepackage{float}
\usepackage{graphicx}
\usepackage{epstopdf}
\usepackage[colorlinks=true,linkcolor=blue,citecolor=blue,urlcolor=blue]{hyperref}
\usepackage[utf8]{inputenc}
\usepackage{indentfirst}
\usepackage{latexsym,bm,euscript}
\usepackage{mathrsfs}
\usepackage{multirow}
\usepackage{natbib}
\usepackage{soul}
\usepackage{subfigure}
\usepackage{times}
\usepackage{tikz}
\usepackage{txfonts}
\usepackage{ulem}
\usepackage{xspace}
\usepackage{xfrac}
\usepackage{physics}
\usetikzlibrary{positioning,calc}

\definecolor{mporange}{RGB}{245, 140, 45}
\definecolor{envgray}{RGB}{205, 210, 218}
\definecolor{projgray}{RGB}{180, 185, 195}
\definecolor{sitegreen}{RGB}{222, 242, 220}
\definecolor{boxblue}{RGB}{46,167,224}
\definecolor{tensororange}{RGB}{237,144,38}
\definecolor{leggray}{RGB}{102,103,103}
\definecolor{ink}{RGB}{36,36,36}

\newcommand{\App}[1]{App.~\ref{#1}}
\newcommand{\Eq}[1]{Eq.~\eqref{#1}}

\def\ket#1{|{#1}\rangle}
\def\bra#1{\langle{#1}|}
\def\dket#1{|{#1}\rangle\!\rangle}
\def\dbra#1{\langle\!\langle{#1}|}

\begin{document}

\title{Perfect Born Sampling of Symmetric Thermal Tensor Network States}

\author{Jianxin Gao}
\thanks{These authors contributed equally to this work.}
\affiliation{Peng Huanwu Collaborative Center for Research and Education, Beihang University, Beijing 100191, China}
\affiliation{Institute of Theoretical Physics, Chinese Academy of Sciences, Beijing 100190, China}

\author{Qiaoyi Li}
\thanks{These authors contributed equally to this work.}
\affiliation{Institute of Theoretical Physics, Chinese Academy of Sciences, Beijing 100190, China}
\affiliation{School of Physical Sciences, University of Chinese Academy of Sciences, Beijing 100049, China}

\author{Yuan Gao}
\thanks{These authors contributed equally to this work.}
\affiliation{Institute of Theoretical Physics, Chinese Academy of Sciences, Beijing 100190, China}

\author{Chuanshu Xu}
\affiliation{Institute of Theoretical Physics, Chinese Academy of Sciences, Beijing 100190, China}
\affiliation{School of Physical Sciences, University of Chinese Academy of Sciences, Beijing 100049, China}

\author{Guoliang Wu}
\affiliation{Institute of Theoretical Physics, Chinese Academy of Sciences, Beijing 100190, China}
\affiliation{School of Physical Sciences, University of Chinese Academy of Sciences, Beijing 100049, China}

\author{Su Yi}
\affiliation{Institute of Fundamental Physics and Quantum Technology, Ningbo University, Ningbo 315211, China}
\affiliation{Peng Huanwu Collaborative Center for Research and Education, Beihang University, Beijing 100191, China}

\author{Bin-Bin Chen}
\affiliation{Peng Huanwu Collaborative Center for Research and Education, Beihang University, Beijing 100191, China}

\author{Wei Li}
\email{w.li@itp.ac.cn}
\affiliation{Institute of Theoretical Physics, Chinese Academy of Sciences, Beijing 100190, China}
\affiliation{Peng Huanwu Collaborative Center for Research and Education, Beihang University, Beijing 100191, China}
\affiliation{Hefei National Laboratory, Hefei 230088, China}

\date{October 13, 2026}
\begin{abstract}
Accurate calculations of correlated fermions at low temperatures constitute a major challenge in quantum many-body physics, most notably in the studies of high-$T_c$ superconductivity and ultracold atom optical lattices. Stochastic sampling of tensor-network states offers a promising route to tackle this problem; however, existing schemes have long faced a fundamental dilemma---sampling efficiency and symmetry acceleration \textit{cannot} be achieved simultaneously. Here we propose a perfect Born sampling approach for thermal tensor networks, which performs importance sampling directly from the purified density matrix via the Born rule and incorporates Abelian and non-Abelian symmetries by sampling symmetry quantum numbers. We benchmark the method, realized as both stochastic matrix product states (stoMPS) and stochastic projected entangled pair states (stoPEPS), on large-scale quantum lattice models. Using stoMPS, we accurately simulate the square-lattice Hubbard model on cylinders up to width $W=10$, and study the triangular-lattice Hubbard model down to $T/t = 1/64$. We reveal scalar chiral order at half filling and kinetic ferromagnetism upon electron doping, with the latter directly compared to ultracold atom experiments. We further extend the stoMPS method to compute finite-temperature quantum dynamics, as demonstrated by the optical conductivity of the Hubbard model. We showcase the generalization of stoPEPS on up to $20\times20$ square-lattice quantum Ising model, where accurate results are obtained near its quantum critical point. Our method combines high sampling efficiency with full symmetry acceleration. The resulting state-of-the-art framework captures both equilibrium and dynamical properties down to ultralow temperatures, thereby providing a powerful theoretical tool for correlated quantum systems, particularly fermion Hubbard quantum gas microscopy.
\end{abstract}
\maketitle

\section{Introduction}
Electron correlation and high-temperature superconductivity pose a prominent open problem in quantum many-body physics~\cite{Lee2006, Keimer2015Nature}, with the Hubbard model and its extensions playing an essential role~\cite{Anderson1987, Arovas2022Review, Qin2022Review}. In recent years, ultracold fermion atoms trapped in optical lattices have emerged as a powerful platform for the quantum simulation of this model, offering single-site-resolved control and readout through quantum gas microscopy~\cite{Boll2016, Gross2017Review, GrossBakr2021, Mazurenko2017, Chiu2019, Hilker2017, Koepsell2019}. Such experiments are advancing rapidly into the intriguing low-temperature regimes of the phase diagram---exploring the pseudogap~\cite{Kendrick2026Pseudogap, Chalopin2026Pseudogap} and strange-metal states~\cite{Brown2019}, while steadily approaching the superconducting regime. Interpreting these experiments relies crucially on accurate finite-temperature theoretical calculations, for calibration and snapshot analysis~\cite{Boll2016, Mazurenko2017}, and ultimately for understanding the correlated many-fermion states and making new predictions. Efficient and controlled low-temperature computational tools for doped large-scale fermion systems, however, remain scarce~\cite{Xu2025Cryogenic}. Current comparisons with quantum gas microscopy, for instance, remain restricted to small clusters or narrow cylinders~\mbox{\cite{Chen2021SLU, Chalopin2026Pseudogap}}.

Tensor networks constitute an efficient and accurate framework for solving quantum many-body problems~\cite{Schollwock2011MPS, Cirac2021RMP}. Beyond ground-state properties, thermal tensor networks have been proposed for simulating finite-$T$ systems~\cite{Bursill1996DMRG, Wang1997, Zwolak2004, Feiguin2005, White2009METTS, Stoudenmire2010, Li2011, Czarnik2012PEPS, METTSvsPurification2015, Dong2017, Chen2017, Chen2018, Li2019, tanTRG2023, Iwaki2021, Goto2021, Iwaki2022, gohlke_thermal_2023, iwaki_sample_2024, Li2026PRBThermal}. In the study of strongly correlated quantum many-body systems, finite-temperature tensor-network calculations play an increasingly important role. Major applications include frustrated quantum spins~\cite{Li2011, Chen2018, Chen2019, Wietek2019PRB, Li2020TMGO, Czarnik2021SS, Gao2022NBCP, HLi2020PRR, YuCPL2021, Li2021NC, Wang2023PRLPlaquette, Gao_double_magnon-roton, gao_seebeck_2025, okubo2025thermalhalltransportkitaev}, correlated electrons~\cite{Czarnik2014fermion, Czarnik2015PEPS, Czarnik2016TNR, Czarnik2019PEPS, Sinha_2022_PRB, Qin2022Review, Wietek2021PRX, Wietek2021PRX-II, Chen2022tbg, Qu2022tJ, tanTRG2023, Qu2023bilayer, Li2026PDW}, and ultracold-atom optical lattices~\cite{Chen2021SLU, Wietek2021PRX, Sinha_2022_PRB}. Such calculations enable the determination of microscopic spin models from measured thermal data of frustrated magnets, give access to exotic low-temperature electronic states in the Hubbard and $t$-$J$ models, and provide quantitative benchmarks for quantum simulations~\cite{Gross2017Review, Mazurenko2017, Hilker2017, Koepsell2021}. Moreover, they can predict novel quantum phases and emergent phenomena, such as spin supersolidity~\cite{Gao2022NBCP, gao_seebeck_2025, Gao_double_magnon-roton, Xiang2024Nature, Shu2026Nature} and high-$T_c$ superconductivity~\cite{tanTRG2023, Qu2022tJ, Qu2023bilayer, Li2026PDW}.

Purification-based thermal tensor-network methods constitute an important approach. The thermal mixed states are represented as ``superstates'' in an enlarged Hilbert space, a construction also known as the thermofield double state~\cite{Takahashi1996TFD}. This approach typically yields a tensor-network representation of the density matrix $\rho(\beta) \equiv e^{-\beta H}/Z(\beta)$, such as a matrix product operator (MPO)~\cite{Feiguin2005, Zwolak2004, Purification2009, Li2011, Dong2017, Chen2017, Chen2018, Li2019, tanTRG2023} or projected entangled pair operator (PEPO)~\cite{Li2011, Czarnik2012PEPS, Czarnik2015PEPS, Czarnik2016TNR, Czarnik2019PEPS}. A distinct advantage of this formulation is its ability to efficiently exploit symmetries~\cite{Chen2018, Li2019, tanTRG2023}, and it delivers highly accurate results in the high- to intermediate-temperature regime. However, the doubled Hilbert space inevitably introduces extra entanglement~\cite{METTSvsPurification2015, Hauschild2018}, which grows rapidly and demands larger bond dimensions at low temperatures.

Stochastic sampling of tensor-network states, exemplified by the minimally entangled typical thermal state (METTS)~\cite{White2009METTS, Stoudenmire2010} and the thermal pure quantum matrix product state (TPQ-MPS)~\cite{Sugiura2012, Iwaki2021} algorithms, offers a promising alternative approach. By acting on tensor-network states rather than operators, it substantially reduces the computational costs~\cite{METTSvsPurification2015, Kusuki2024}, due to the fewer physical indices and potentially lower entanglement. Nevertheless, the performance of stochastic methods rests on the sampling scheme: how the samples are generated, whether each sample preserves the symmetry, and how the thermal weights enter the ensemble average. For instance, importance sampling has been realized in the METTS algorithm, where the thermalized Markov chain carries the thermal weights through the sampling frequency. Yet, existing schemes have long struggled to reconcile sampling efficiency with symmetry acceleration~\cite{Bruognolo2015, Binder2017SYMETTS}, constraining the accessible system sizes and temperatures. Developing efficient and accurate thermal tensor-network methods for large-scale two-dimensional (2D) Hubbard lattices remains an open challenge.

To this end, we introduce perfect Born sampling of thermal tensor networks, a highly efficient approach realized as stochastic matrix product states (stoMPS) and stochastic projected entangled pair states (stoPEPS). The framework takes inspiration from the experimental snapshots of quantum gas microscopy~\cite{Gross2017Review, Mazurenko2017, GrossBakr2021}. A Born sample $\ket{\psi_\alpha}$ is obtained by drawing a product configuration $\alpha$ of the auxiliary space according to the Born rule and projecting the purified thermal supervector onto it. The ensemble of Born samples thus provides a faithful statistical representation of the thermal state, without Markov chains or autocorrelation---in this sense, the sampling is \textit{perfect}. Moreover, Abelian and non-Abelian symmetries can be incorporated directly into the sampling, which naturally accommodates the spin and charge symmetries of the Hubbard model and substantially improves the computational efficiency. The approach achieves high accuracy and computational speed by combining importance sampling with full symmetry acceleration, which has long eluded existing stochastic schemes. With statistically independent samples evolved in parallel, we simulate square-lattice Hubbard cylinders up to the unprecedented width~$W=10$ and quantum Ising lattices of size $20\times20$.

The remainder of this paper is organized as follows.
Section~\ref{Sec:Alg} describes the Born sampling algorithm, with particular emphasis on the symmetry-based sampling scheme.
Section~\ref{Sec:SLHubbard} benchmarks the method on the half-filled square-lattice Hubbard model against purification and quantum Monte Carlo.
Section~\ref{Sec:app} applies it to the triangular-lattice Hubbard model, addressing scalar chiral spin order at half filling (Sec.~\ref{Sec:chirality}) and kinetic ferromagnetism upon electron doping (Sec.~\ref{Sec:kinetic_ferromagnetism}).
Section~\ref{Sec:conductivity} presents the optical conductivity of the Hubbard model as a demonstration of finite-temperature dynamical calculations, and Sec.~\ref{Sec:stoPEPS} generalizes the Born sampling to stoPEPS with an application to the 2D quantum Ising model. Section~\ref{Sec:Discussion} further discusses the connections to METTS and TPQ-MPS approaches, as well as the purification basis of the framework.
Additional benchmarks and technical details are collected in the Appendices, including the
pseudocode and construction of the Born sampling
algorithm (App.~\ref{App:algorithm}),
the annealed Born resampling scheme (App.~\ref{App:resample}),
the imaginary-time evolution and computational costs (App.~\ref{App:CBE-TDVP}),
canonical-ensemble sampling at fixed particle number (App.~\ref{App:fixed_particle}), and the details of stoPEPS algorithm (App.~\ref{App:fu-stoPEPS}).

\section{Born-sampling thermal tensor networks}
\label{Sec:Alg}

\begin{figure*}[t]
\includegraphics[width=\linewidth]{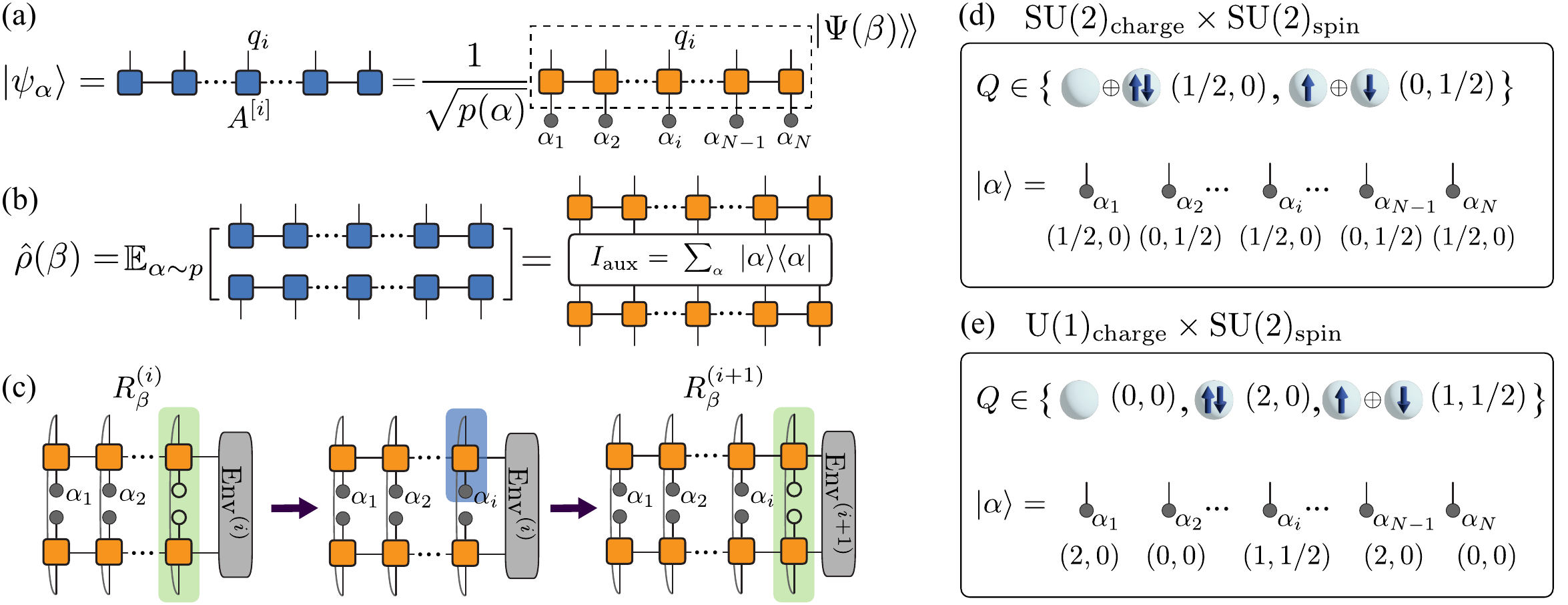}
\caption{Diagrammatic representation of stoMPS method.
(a) A normalized Born sample $\ket{\psi_\alpha}$ is obtained by projecting the auxiliary indices of the purified supervector $\dket{\Psi(\beta)}$ (orange tensors) onto a product configuration $\alpha = (\alpha_1, \dots, \alpha_N)$ [cf. Eq.~\eqref{Eq:Born}]. A sampled MPS is written as $\ket{\psi} = \sum_{q_1, \dots, q_N} \left( A_{q_1}^{[1]} A_{q_2}^{[2]} \dots A_{q_N}^{[N]} \right) \ket{q_1, q_2, \dots, q_N}$, where $q_i$ denotes the local physical state at site $i$, and $A_{q_i}^{[i]}$ is a matrix whose successive product yields the coefficient. Each tensor $A^{[i]}$ is shown as a blue solid node, with physical index $q_i$ (vertical) and virtual indices (horizontal) connecting adjacent sites.
(b) The unbiasedness condition in Eq.~\eqref{Eq:unbiased}: contracting the auxiliary indices between the supervector $\dket{\Psi(\beta)}$ and its conjugate $\dbra{\Psi(\beta)}$ yields the thermal density matrix $\rho(\beta)$, and inserting the resolution of the auxiliary identity $I_{\rm aux} = \sum_{\alpha} \ket{\alpha}\bra{\alpha}$ regroups it into the ensemble average $\mathbb{E}[\ket{\psi_\alpha}\bra{\psi_\alpha}]$ over Born samples (blue MPSs).
(c) Sequential Born sampling of the auxiliary configuration via the chain rule.
With the preceding indices $(\alpha_1, \dots, \alpha_{i-1})$ fixed (filled circles), contracting the doubled network,
including the two local MPO tensors at site $i$ and the gray right-environment tensor ${\rm Env}^{(i)}$,
while leaving the two auxiliary indices at site $i$ open, and normalizing it yields the local conditional reduced density matrix $R_{\beta}^{(i)}$.
The green shading marks the current sampling site.
The diagonal entries of $R_{\beta}^{(i)}$ give the conditional probability $p(\alpha_i \mid \alpha_1, \dots, \alpha_{i-1})$  (see App.~\ref{App:algorithm}).
Once $\alpha_i$ is sampled, projecting the auxiliary leg of the local MPO tensor onto $\bra{\alpha_i}$ yields the $i$th tensor of the sampled MPS (blue shade). The sampling center then advances to site $i+1$, and the procedure repeats until a complete configuration $\alpha$ is obtained.
(d, e) Local Hilbert-space bases under different symmetry groups, where the quantum numbers $Q = (C, S)$ denote charge and spin, respectively~\cite{Weichselbaum2012, weichselbaum2024QSpace, weichselbaum2024QSpace_code, Devos2025TensorKit}.
(d) Under ${\rm SU(2)_{charge}} \times {\rm SU(2)_{spin}}$ symmetry, each local site possesses two basis multiplets, labeled by $(C=1/2, S=0)$ and $(C=0, S=1/2)$.
(e) Under ${\rm U(1)_{charge}} \times {\rm SU(2)_{spin}}$ symmetry, each local site instead possesses three basis sectors, labeled by $(C_z=0, S=0)$, $(C_z=2, S=0)$, and $(C_z=1, S=1/2)$.
}
\label{Fig1}
\end{figure*}

\subsection{Born sampling of stoMPS}
Stochastic sampling methods offer a powerful approach for calculating finite-temperature properties of quantum many-body systems. The thermal density matrix $\rho(\beta) \equiv e^{-\beta H}/Z(\beta)$, with $Z(\beta) = \mathrm{Tr}\, e^{-\beta H}$ the partition function, can be decomposed into an ensemble of pure states drawn directly from the thermal distribution. To construct the Born sampling scheme, we recast the half-evolved density operator as a purified supervector in the doubled Hilbert space,
\begin{equation}
\dket{\Psi(\beta)} \equiv \frac{e^{-\beta H/2}}{\sqrt{Z(\beta)}}\dket{I},
\label{Eq:R}
\end{equation}
where $\dket{X} \equiv \sum_{i,j} X_{ij} \ket{i} \otimes \ket{j}$ denotes the vectorization of an operator $X$ into the physical$\,\otimes\,$auxiliary Hilbert space, as in the thermofield double construction~\cite{Takahashi1996TFD}.

Taking a 1D quantum system as an example, the supervector admits a matrix-product representation,
\begin{equation}
\begin{aligned}
\dket{\Psi(\beta)} ={}& \sum_{\{q_i\},\{\alpha_i\}} \left( A_{q_1,\alpha_1}^{[1]} A_{q_2,\alpha_2}^{[2]} \dots A_{q_N,\alpha_N}^{[N]} \right) \\
& \times \ket{q_1, \dots, q_N} \otimes \ket{\alpha_1, \dots, \alpha_N},
\end{aligned}
\label{Eq:superMPS}
\end{equation}
where each local tensor $A_{q_i,\alpha_i}^{[i]}$ carries a physical index $q_i$ and an auxiliary index $\alpha_i$, and their successive matrix product over the virtual bonds yields the coefficient. In practice, we represent the ``super-MPS'' in the doubled Hilbert space equivalently as an MPO, with $q_i$ and $\alpha_i$ playing the roles of the output and input indices, respectively.

As shown in Fig.~\ref{Fig1}(a), we choose a complete orthonormal product basis $\{\ket{\alpha}\}$ of the auxiliary space, with $\alpha = (\alpha_1, \dots, \alpha_N)$ for $N$ lattice sites. We then apply the Born rule to the auxiliary indices and draw a configuration $\alpha$ according to its Born probability $p(\alpha)$. Projecting the supervector onto $\bra{\alpha}$ then yields the normalized pure state
\begin{equation}
\ket{\psi_\alpha} = \frac{\langle \alpha | \Psi(\beta) \rangle\!\rangle}{\sqrt{p(\alpha)}},
\qquad
p(\alpha) = \| \langle \alpha | \Psi(\beta)\rangle\!\rangle \|^2,
\label{Eq:Born}
\end{equation}
with normalized probabilities $\sum_\alpha p(\alpha) = 1$. The resulting collapsed state $\ket{\psi_\alpha}$ is referred to as a Born sample throughout, sharing the same probability $p(\alpha)$. As demonstrated in Fig.~\ref{Fig1}(b), the ensemble of sampled states reproduces the thermal density matrix exactly:
\begin{equation}
\mathbb{E}\left[ \ket{\psi_\alpha}\bra{\psi_\alpha} \right]
= \sum_\alpha p(\alpha)\, \frac{\langle \alpha | \Psi(\beta) \rangle\!\rangle \langle\!\langle \Psi(\beta) | \alpha \rangle}{p(\alpha)}
= \rho(\beta),
\label{Eq:unbiased}
\end{equation}
where $\mathbb{E}[\cdots] \equiv \sum_{\alpha} p(\alpha)\,[\cdots]$ denotes the expectation with $\alpha$ sampled from Born probability $p(\alpha)$ and $\mathrm{Tr}_{\rm aux}[\dket{\Psi(\beta)}\dbra{\Psi(\beta)}] = \rho(\beta)$. Consequently, the thermal expectation value of an observable $O$ is given by
\begin{equation}
\langle O \rangle_\beta = \mathrm{Tr}[\rho(\beta) O] = \mathbb{E}\left[ \bra{\psi_\alpha} O \ket{\psi_\alpha} \right],
\label{Eq:Observable}
\end{equation}
estimated by the plain average $\langle O \rangle_\beta \simeq \frac{1}{N_{\rm s}} \sum_{k=1}^{N_{\rm s}} \bra{\psi_{k}} O \ket{\psi_{k}}$ over $N_{\rm s}$ Born samples $\ket{\psi_{k}} \equiv \ket{\psi_{\alpha(k)}}$.

We term this sampling scheme \textit{perfect Born sampling}: the Born probability is already encoded in the sample frequency, such that each configuration is sampled independently and distributed exactly according to the Boltzmann thermal distribution, without thermalization or reweighting. Such direct sampling from the Born probability of a tensor-network state can be traced back to the perfect sampling of tensor networks~\cite{Ferris2012, Vieijra2021}, and has also been exploited for generative modeling in machine learning, where it is known as direct sampling of a Born machine~\cite{Han2018PRX}.

In practice, the sampling starts from an intermediate temperature $T_0\equiv1/\beta_0$, at which the supervector $\dket{\Psi(\beta_0)}$ is prepared, e.g., by tanTRG~\cite{tanTRG2023, Li2026PRBThermal}.
The Born sampling is then performed sequentially via the chain rule
$p(\alpha_1, \dots, \alpha_N) = p(\alpha_1) \prod_{i=2}^{N} p(\alpha_i | \alpha_1, \dots, \alpha_{i-1})$.
At each site $i$, the conditional probability $p(\alpha_i | \alpha_1, \dots, \alpha_{i-1})$ of the local auxiliary index $\alpha_i$ is evaluated from the local conditional reduced density matrix $R_{\beta}^{(i)}$ shown in Fig.~\ref{Fig1}(c) and App.~\ref{App:algorithm}.
The local auxiliary index is then sampled according to this conditional probability, and the corresponding local tensor of the Born sample $\ket{\psi_\alpha}$ is obtained by projecting the local MPO tensor onto the sampled auxiliary state. The sampling center then advances to site $i+1$, and the procedure repeats until a complete configuration $\alpha$ is obtained. The construction is thus hybrid by design: the high- to intermediate-temperature regime is handled by purification~\cite{Li2011, Chen2018, tanTRG2023}, which is accurate and efficient and in turn provides a high-precision parent distribution for the sampling, while the low-temperature regime is covered by evolving MPS samples, whose much lower entanglement than the full density-matrix MPO yields both higher accuracy at a given bond dimension and significantly lower computational cost.

Upon further cooling, each Born sample follows
\begin{equation}
\ket{\tilde{\psi}_\alpha(\beta)} = e^{-(\beta-\beta_0)H/2} \ket{\psi_\alpha(\beta_0)},
\label{Eq:continue}
\end{equation}
where the evolved sample is no longer normalized, and its squared norm records the thermal weight accumulated during the additional evolution from $\beta_0$ to $\beta$. The imaginary-time evolution is performed with the one-site time-dependent variational principle (TDVP) algorithm~\cite{TDVP2011, TDVP2016} with controlled bond expansion (CBE)~\cite{Gleis2023CBE, JhengWeionetdvp}, as detailed in App.~\ref{App:CBE-TDVP}. The thermal expectation value and the partition function then follow from
\begin{equation}
\langle O \rangle_\beta = \frac{\mathbb{E}[ \bra{\tilde{\psi}_\alpha(\beta)} O \ket{\tilde{\psi}_\alpha(\beta)}]}{\mathbb{E} [\langle \tilde{\psi}_\alpha(\beta)|\tilde{\psi}_\alpha(\beta)\rangle]},
\label{Eq:ratio}
\end{equation}
and
\begin{equation}
\ln Z(\beta) = \ln Z(\beta_0) + \ln \mathbb{E}
\left[ \| \tilde{\psi}_\alpha(\beta) \|^2 \right],
\label{Eq:lnZ}
\end{equation}
respectively. The detailed workflow of Born sampling algorithm is summarized in Alg.~\ref{alg:QMPS} in App.~\ref{App:algorithm}.

As the cooling interval $(\beta-\beta_0)$ grows, however, the estimator in Eq.~\eqref{Eq:ratio}, albeit exact, gradually loses efficiency: the squared norms of the cooled samples spread over an ever wider range, and the effective sample size decreases. To restore the sampling efficiency, we devise an annealed Born resampling scheme: whenever the sampling efficiency falls below a prescribed threshold, fresh auxiliary configurations are drawn from a proposal distribution constructed from the cooled ensemble and projected onto the supervector at $\beta_0$. The details can be found in App.~\ref{App:resample}, where annealed resampling is shown to restore nearly uniform sample norms and thus the effective sample size.

\subsection{Symmetric Born sampling}
Abelian and non-Abelian symmetries can be explicitly incorporated in the Born sampling framework. Such symmetries can be exploited in the MPO or PEPO representations~\cite{weichselbaum2024QSpace, weichselbaum2024QSpace_code, Devos2025TensorKit, Liu2015Simplex} of the thermal density matrix (or equivalently the supervector $\dket{\Psi(\beta)}$), as demonstrated in the exponential tensor renormalization group (XTRG)~\cite{Chen2018} and tanTRG~\cite{tanTRG2023} methods. Crucially, projecting the auxiliary indices of a symmetric supervector onto symmetry multiplets yields Born samples that are themselves symmetry eigenstates, i.e., the sampled MPSs preserve the symmetry. Consequently, the same symmetric block structure accelerates both the sampling process and the subsequent evolution of each Born sample.

As illustrated in Fig.~\ref{Fig1}(d,e), the local Hilbert space of a fermionic site consists of four fundamental basis states: the empty state $\ket{0}$, the spin-up state $\ket{\uparrow}$, the spin-down state $\ket{\downarrow}$, and the doubly occupied state $\ket{\uparrow\downarrow}$. Each symmetry basis state carries an irreducible representation of either the Abelian $\mathrm{U(1)}$ or non-Abelian $\mathrm{SU(2)}$ symmetry group and is labeled by a quantum number $Q$ associated with the conserved charge $C$ and spin $S$. These symmetry eigenstates play the role of the product states $\ket{\alpha}$ in the Born sampling: following the sampling procedure illustrated in Fig.~\ref{Fig1}(a,c), the eigenstates are sampled and the supervector is collapsed onto them, yielding symmetric MPS samples. Sampling in our method is thus performed directly within these symmetric local bases.

For SU(2) particle-hole and spin-rotation symmetries, denoted as $\mathrm{SU(2)_{charge} \times SU(2)_{spin}}$, the local Hilbert space is organized as the direct product of charge and spin irreducible representations. The empty and doubly occupied states, $\ket{0}$ and $\ket{\uparrow\downarrow}$, together form a charge-$1/2$ doublet with total spin $S = 0$, while the singly occupied states, $\ket{\uparrow}$ and $\ket{\downarrow}$, form a charge singlet with spin $S = 1/2$. Each basis state is therefore labeled by the composite quantum numbers $(C, S)$ from the charge and spin sectors~\cite{Weichselbaum2012, weichselbaum2024QSpace, weichselbaum2024QSpace_code, Devos2025TensorKit}. We perform Born sampling in the multiplets labeled by $Q=(C,S)\in\{(1/2,0),(0,1/2)\}$, as illustrated in Fig.~\ref{Fig1}(d). For Abelian U(1) charge conservation and spin-rotation SU(2) symmetry, denoted as $\mathrm{U(1)_{charge} \times SU(2)_{spin}}$, the local Hilbert space is labeled by the charge number $C_z$ and spin quantum number $S$. As shown in Fig.~\ref{Fig1}(e), the basis states are $\ket{0}$ with $(C_z, S) = (0, 0)$, $\ket{\uparrow}$ and $\ket{\downarrow}$ with $(C_z, S) = (1, 1/2)$, and $\ket{\uparrow\downarrow}$ with $(C_z, S) = (2, 0)$.

The same symmetry-guided sampling procedure applies to PEPO representations of the supervector, where the sampled symmetry eigenstates collapse the PEPO into symmetric PEPS samples. The annealed resampling (see App.~\ref{App:resample}) carries over to the symmetric setting without modification, as the regeneration projections employ the same symmetric basis.

\begin{figure*}[t]
\includegraphics[width=\linewidth]{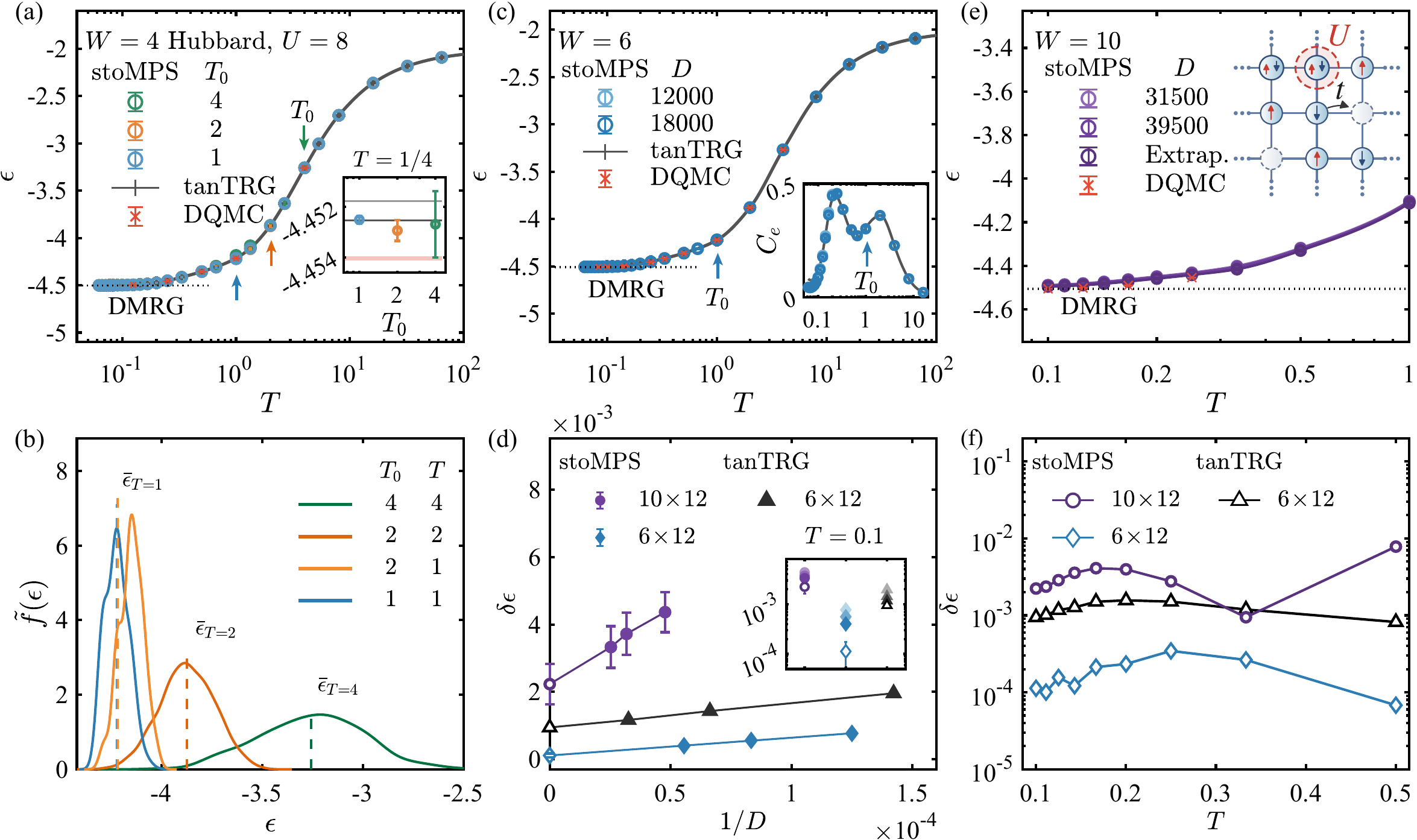}
\caption{Finite-temperature benchmarks of stoMPS on Hubbard cylinders up to width $W=10$.
(a) Energy per site $\epsilon$ on CL $4\times8$ cylinder, obtained with $N_{\rm s}=400$ Born samples {with $D = 12000$}
taken at $T_0/t=4$, $2$, and $1$. The  tanTRG (gray line, $D = 24000$), DQMC (red crosses), and DMRG ground-state energy (horizontal dashed line) are shown for comparison. Inset shows the three stoMPS estimates at $T=1/4$ against the DQMC reference with its error range (red band) and the tanTRG results for $D=12000$ and $24000$ (light and dark gray lines, respectively). The stoMPS relative errors are $\lesssim 3.4\times10^{-4}$.
(b) Unweighted sample-energy distributions $\tilde f(\epsilon)$ for the $4\times8$ calculations, with the Born samples taken at $T_0$ and the distributions measured at $T$. Solid curves are kernel-density estimates, and dashed vertical lines mark the corresponding average energies.
(c) Results on CL $6\times12$ cylinder from $N_{\rm s}=200$ Born samples initialized at $T_0=1$, and the tanTRG results with extrapolation are taken from Ref.~\cite{tanTRG2023}. The inset shows the specific heat $C_e$, with Born sampling temperature $T_0$ located between the two peaks.
(d) Relative deviation $\delta\epsilon$ against DQMC at $T=0.1$, where filled (open) symbols show finite-$D$ (extrapolation) results. The inset displays $\delta\epsilon$ on a logarithmic vertical scale.
(e) $\epsilon$ on CL $10\times12$ cylinder from $N_{\rm s}=100$ Born samples with $D\simeq31500$ and $39500$, respectively. The DQMC and the DMRG results are shown for reference.
(f) $\delta\epsilon$ for the extrapolated stoMPS results on the CL $10\times12$ and $6\times12$ cylinders, reaching $2.2\times10^{-3}$ and $1.1\times10^{-4}$, respectively, at $T=0.1$; the width-6 errors remain well below the extrapolated tanTRG results on the same cylinder throughout the plotted temperature range.
}
\label{Fig2}
\end{figure*}

\section{Square-lattice Hubbard Model}
\label{Sec:SLHubbard}
We benchmark the stoMPS method on the square-lattice Hubbard model at half filling,
\begin{equation}
H = -t \sum_{\langle i,j \rangle,\sigma} \left(c_{i\sigma}^\dagger c^{\phantom{\dagger}}_{j\sigma} + \text{h.c.}\right) + U \sum_i n_{i\uparrow} n_{i\downarrow} -\mu\sum_{i,\sigma} n_{i\sigma},
\label{Fermi-Hubbard}
\end{equation}
where $ c_{i\sigma}^\dagger $ and $ c^{\phantom{\dagger}}_{i\sigma} $ create and annihilate a fermion at site $i$ with spin $\sigma \in \{ \uparrow, \downarrow \}$, and $n_{i\sigma} = c_{i\sigma}^\dagger c^{\phantom{\dagger}}_{i\sigma}$ is the corresponding number operator.
We set the nearest-neighbor hopping amplitude $t=1$ as the energy unit and the interaction strength $U=8$. When chemical potential is set as $\mu=U/2$, half filling is enforced and particle-hole symmetry allows us to exploit the full $\mathrm{SU(2)}_{\rm charge} \times \mathrm{SU(2)}_{\rm spin}$ symmetry. We consider cylindrical geometries CL $W\times L$, with circumference $W$ and length $L$, and benchmark against determinant quantum Monte Carlo (DQMC)~\cite{Blankenbecler1981, Assaad2008}, the purification tanTRG~\cite{tanTRG2023}, and ground-state density matrix renormalization group (DMRG) calculation, whose bond dimensions are chosen such that the truncation errors are of the order of $10^{-5}$.

Figure~\ref{Fig2}(a) shows the energy per site, $\epsilon=E/N$, of the half-filled Hubbard model on the CL $4\times8$ cylinder.  We compare calculations with Born sampling initiated at different temperatures $T_0 \equiv 1/\beta_0=4$, $2$, and $1$. The three estimates are mutually consistent and closely follow DQMC benchmarks throughout. At $T=1/4$, all three stoMPS energy estimates are closer to the DQMC reference than the tanTRG result at the same bond dimension. Thus, all Born sampling temperatures considered yield accurate results: the relative deviations of the $D=12000$ stoMPS results from the numerically exact DQMC reference lie below those of the tanTRG result with $D=24000$. This accuracy is achieved at a reduced computational cost: the mean evolution time (from $T_0$ to the low temperature $T=1/16$) per stoMPS sample is one tenth of the tanTRG runtime (see App.~\ref{App:CBE-TDVP}). Moreover, lowering $T_0$ reduces the statistical error bars, see the inset of Fig.~\ref{Fig2}(a).

This mechanism is illustrated more directly in Fig.~\ref{Fig2}(b), which compares the distributions of the per-site sample energies at different sampling ($T_0$) and observation ($T$) temperatures. Here, $\tilde{f}(\epsilon)$ denotes the empirical energy distribution obtained by kernel-density estimation. The dashed vertical lines mark the corresponding thermal-average energies, with the continuation weights applied for $T<T_0$. At $T_0$, the configurations are generated by exact sequential Born sampling from the purified thermal state. After continuation to a lower temperature, the ensemble generated at $T_0=2$ and evolved to $T=1$ retains substantial overlap with the direct Born sampling distribution at $T_0=1$. Note that at ultralow temperatures, where the overlap becomes small, the annealed Born resampling scheme can be employed to restore the perfect Born distribution and hence the sampling efficiency (see App.~\ref{App:resample}).

Figure~\ref{Fig2}(c) extends the calculations to the CL $6\times12$ cylinder, with Born sampling initiated at $T_0=1$ and several bond dimensions $D$. As shown in the inset of Fig.~\ref{Fig2}(c), $T_0=1$ lies just below the charge peak of electronic specific heat, serving as a proper intermediate temperature for the Born sampling.
The stoMPS results accurately reproduce the temperature dependence of $\epsilon$, in close agreement with the benchmark DQMC and tanTRG data. At the lowest temperature $T=0.1$, the relative deviation of stoMPS results from the reference DQMC data, defined as $\delta\epsilon\equiv |\epsilon-\epsilon_{\rm DQMC}|/|\epsilon_{\rm DQMC}|$, decreases systematically with increasing $D$.

In Fig.~\ref{Fig2}(d), we perform a linear extrapolation of $\epsilon$ in $1/D$, with the corresponding tanTRG data included for comparison. The extrapolated results on the width-6 and width-10 cylinders are shown in Fig.~\ref{Fig2}(f). For the width-6 cylinder, the stoMPS deviations remain at the level of a few $10^{-4}$ across the whole range, achieving an order-of-magnitude improvement over tanTRG. For this width-10 cylinder, the extrapolated deviation remains at the sub-percent level throughout the plotted range and reaches a few $10^{-3}$ at the lowest temperature. Together, the DQMC comparisons across all three cylinder widths demonstrate the accuracy and scalability of stoMPS for finite-temperature simulations.

\section{Triangular-lattice Hubbard model}
\label{Sec:app}
Below we investigate the triangular-lattice Hubbard model with stoMPS, examining chiral order at half filling and the emergence of kinetic ferromagnetism upon doping. These problems remain highly challenging for Monte Carlo methods owing to the notorious sign problem. Moreover, stoMPS enables efficient canonical-ensemble simulations with a fixed particle number (see App.~\ref{App:fixed_particle}), suited for the doped Hubbard systems.

\subsection{Chiral order at half filling}
\label{Sec:chirality}
Previous studies have revealed a rich phase structure in the half-filled triangular-lattice Hubbard model. For a cylinder with width $W = 4$, three distinct phases have been identified: a metallic phase in the weak-coupling regime ($U/t \lesssim 9$), a 120$^{\circ}$ Heisenberg antiferromagnetic (AFM) phase in the strong-coupling limit ($U/t \gtrsim 10.75$), and an intermediate chiral spin liquid (CSL) phase ($9 \lesssim U/t \lesssim 10.75$)~\cite{szasz_chiral_2020, binbinChen-chiral, zhu_chiral_2024}.

\begin{figure}[t]
\includegraphics[width=1\linewidth]{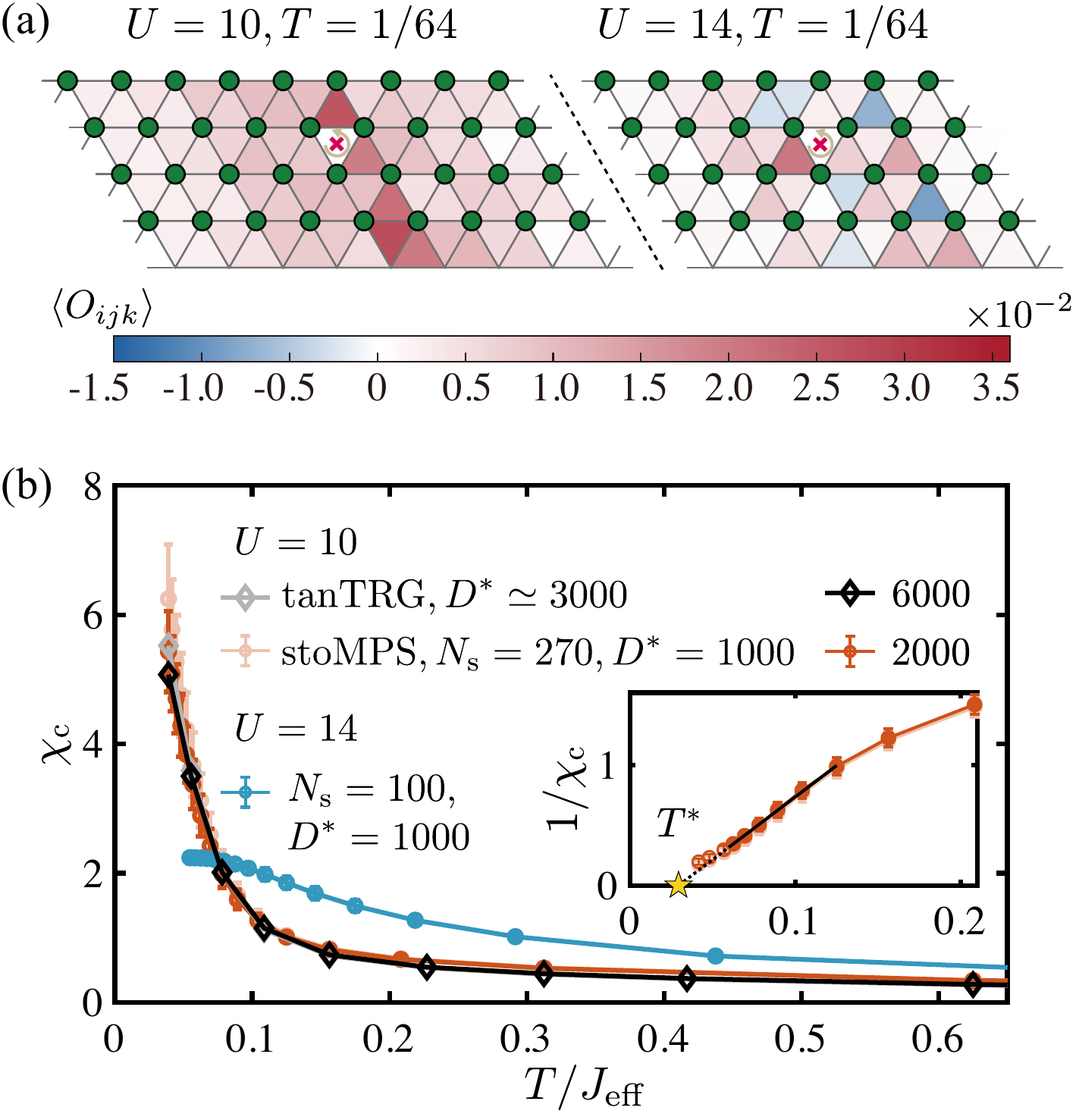}
\caption{
Chiral response of the half-filled triangular-lattice Hubbard model.
(a) Local chirality $\langle O_{ijk} \rangle$ induced by a pinning field ($h_c = 0.1$) applied to the central triangle (red cross) on a YC $4\times12$ cylinder at low temperature ($T = 1/64$). Results are shown for $U = 10$ (CSL phase) and $U = 14$ (120$^\circ$ AFM phase).
(b) Temperature dependence of the chiral susceptibility $\chi_{\rm c}$, where the effective exchange interaction
$J_{\rm eff} \equiv 4t^2/U$. The stoMPS results at $U=10$ (light- and dark-orange lines) are benchmarked against tanTRG calculations with $D^*\simeq3000$ (gray line) and $6000$ (black line). The inset shows $1/\chi_{\rm c}$ versus $T/J_{\rm eff}$. A linear fit over the range $0.06 \leq T/J_{\rm eff} \leq 0.125$ yields an extrapolated characteristic temperature of chiral fluctuations, $T^*/J_{\rm eff} \simeq 0.03$.
}
\label{Fig:chiral}
\end{figure}

The scalar spin chirality operator, defined as $ O_{ijk} \equiv (\bm{S}_i \times \bm{S}_j) \cdot \bm{S}_k $, measures the scalar chirality of three neighboring spins at the vertices of an elementary triangle. To probe the chiral properties of the system, we introduce a small local pinning field into the Hamiltonian as follows:
\begin{equation}
    H = -t \sum_{\langle i,j \rangle,\sigma} \left(c_{i\sigma}^\dagger c^{\phantom{\dagger}}_{j\sigma} + \text{h.c.}\right) + U\sum_i n_{i\uparrow}n_{i\downarrow} - h_{c} (\bm{S}_{i_0} \times \bm{S}_{j_0}) \cdot \bm{S}_{k_0},
    \label{eq:chiral_hamiltonian}
\end{equation}
where $i_0, j_0, k_0$ index the pinned triangle, as illustrated in Fig.~\ref{Fig:chiral}(a), and $h_c$ is the strength of the chiral pinning field. As the pinning field retains spin SU(2) symmetry, we exploit ${\rm U(1)_{charge} \times SU(2)_{spin}}$ symmetry in the calculations.

\begin{figure}[t]
\includegraphics[width=1\linewidth]{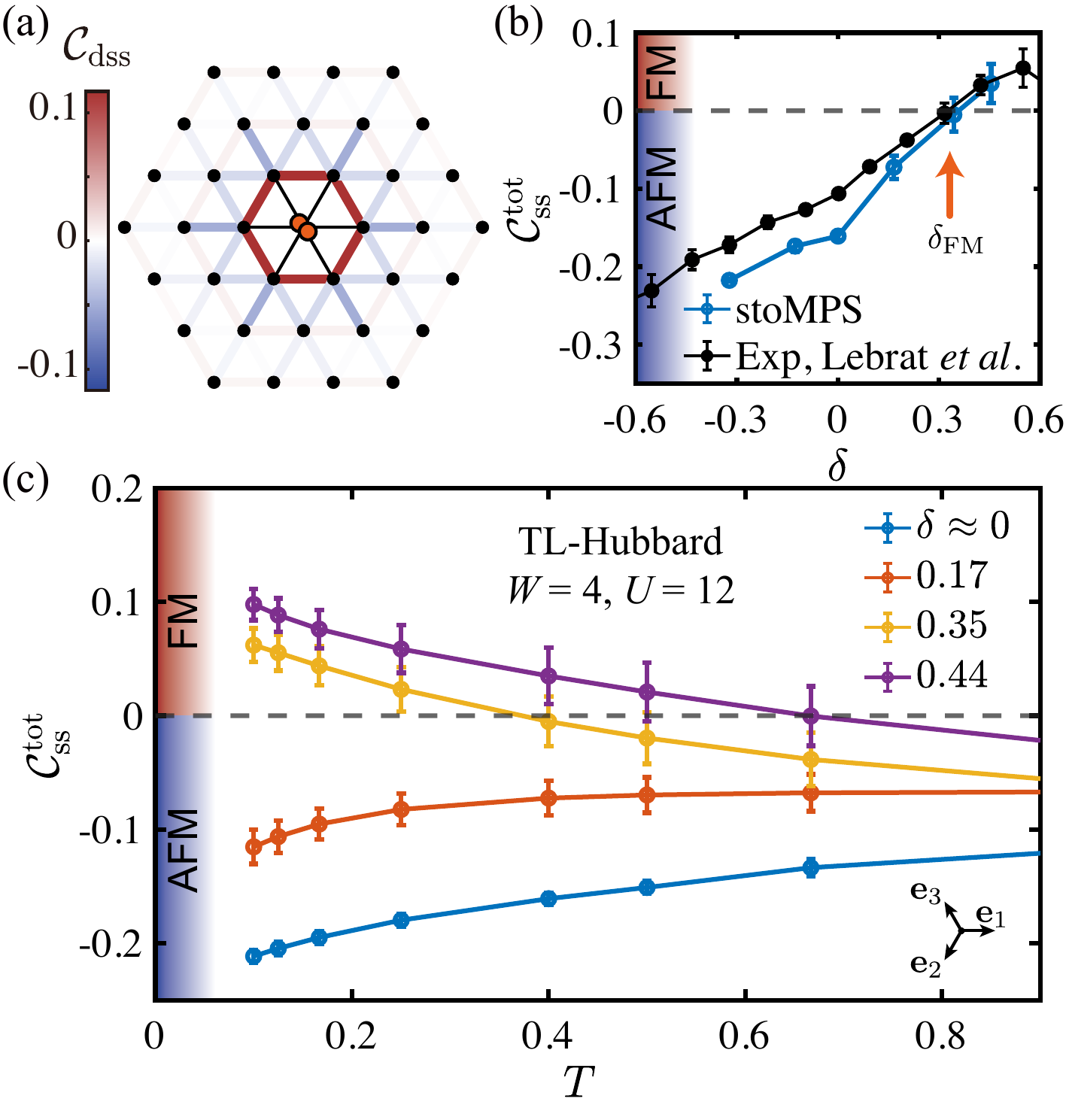}
\caption{
Kinetic ferromagnetism in the electron-doped triangular-lattice Hubbard model.
(a) $\mathcal{C}_{\rm dss}$ calculated on a finite cluster with $U = 12$, $T=1/16$, and electron doping $\delta = 0.16$. The data are symmetrized according to the $C_6$ rotational symmetry. Red (blue) indicates FM (AFM) correlations.
(b) Bulk-averaged nearest-neighbor spin correlation $\mathcal{C}_{\rm ss}^{\rm tot}$ versus doping $\delta$. The stoMPS results are obtained on a YC $4\times12$ cylinder at $U = 12$, $T = 0.4$, compared to experimental data from Ref.~\cite{lebrat_observation_2024} [$U = 9.9(3)$, $T = 0.62(4)$].
(c) Temperature evolution of $\mathcal{C}_{\rm ss}^{\rm tot}$ at
fixed total particle numbers $N_{\rm e} = 48$,
$56$, $64$, and $68$. The average bulk doping
levels $\delta$ are indicated in the legend. The lower right corner indicates the three directions $\bm{\mathrm{e}}_\nu$ ($\nu=1,2,3$).
The stoMPS simulations exploit $\mathrm{U(1)}_{\rm charge} \times \mathrm{SU(2)}_{\rm spin}$ symmetry with a bond dimension $D^*=1000$ in (a) and $D^*=1500$ in (b) and (c), and $N_{\rm s}=100$ samples throughout.
}
\label{Fig:FM_TL_Hubbard}
\end{figure}

We perform stoMPS simulations
on a $4\times12$ Y-cylinder (YC) Hubbard model,
focusing on the CSL phase at $U = 10$ and the 120$^{\circ}$ AFM phase at $U = 14$. In Fig.~\ref{Fig:chiral}(a), we show the spatial distribution of chirality $\langle O_{ijk}\rangle$ at low temperature. In the CSL phase, we observe a prominent chiral response in the bulk region, reflecting strong chiral correlations. When the interaction strength $U$ is increased to 14, the chiral response becomes weak, featuring alternating positive and negative regions in stark contrast to the CSL state. This behavior indicates that the induced chirality is dominated by local fluctuations, consistent with the transition out of the CSL phase into the 120$^{\circ}$ AFM phase as $U$ increases.

We define the chiral susceptibility as $ \chi_{\rm c} =  \sum_{\langle ijk \rangle} \langle O_{ijk} \rangle/h_c $, which is shown in Fig.~\ref{Fig:chiral}(b).
At $ U = 10 $, the $ \chi_{\rm c} $ calculated by stoMPS exhibits divergent behavior at low temperature, consistent with the results obtained from tanTRG. This divergent behavior is indicative of the development of long-range chiral correlations in the low-temperature CSL phase. To further investigate this, we show $ 1/\chi_{\rm c} $ versus
$ T/J_{\rm eff} $ in the inset, together with a linear fit at intermediate temperatures, with the temperature measured in units of the effective spin exchange $J_{\rm eff} \equiv 4t^2/U$. By extrapolating this fit to $1/\chi_{\rm c} = 0$, we estimate the characteristic ordering temperature scale as $T^*/J_{\rm eff} \simeq 0.03$. In contrast, at $ U = 14 $, the chiral susceptibility shows no such divergence at low temperatures, as the system is in the 120$^{\circ}$ AFM phase at this interaction strength. These finite-temperature calculations of chiral response functions clearly support the quantum phase transition between the CSL and 120$^{\circ}$ AFM phases in the half-filled triangular-lattice Hubbard model.

\subsection{Kinetic ferromagnetism at finite doping}
\label{Sec:kinetic_ferromagnetism}

Kinetic ferromagnetism in the doped Hubbard model has recently attracted great interest, particularly on the triangular lattice where geometric frustration and particle-hole asymmetry lead to rich magnetic phase diagrams~\cite{nagaoka_ferromagnetism_1966, tasaki_extension_1989, haerter_kinetic_2005}. While hole doping typically favors antiferromagnetic correlations due to quantum interference effects, electron doping is predicted to stabilize ferromagnetic correlations to gain kinetic energy~\cite{xu_frustration_2023, lebrat_observation_2024}. Recent quantum gas experiments have provided compelling evidence for such Nagaoka polarons~\cite{xu_frustration_2023, lebrat_observation_2024, morera_high-temperature_2023, chen_ferromagnetism_2024}. Previous numerical studies using DQMC and the finite-temperature Lanczos method (FTLM) have indeed confirmed the emergence of ferromagnetic correlations in the electron-doped regime at strong coupling~\cite{lebrat_observation_2024, lee_triangular_2023}. However, reaching sufficiently low temperatures remains challenging for these methods: DQMC is hampered by the sign problem, while FTLM is restricted to small system sizes.

Here, stoMPS offers a powerful method for exploring kinetic ferromagnetism in the doped triangular-lattice Hubbard model, enabling us to provide numerical benchmarks and elucidate the microscopic mechanisms underlying recent experimental observations. To probe Nagaoka polarons induced by electron doping, we evaluate the connected three-point correlation function introduced in Ref.~\cite{lebrat_observation_2024}:
\begin{equation}
\begin{aligned}
    \mathcal{C}_{\rm dss}(\bm{\mathrm{r}}_0;
    \bm{\mathrm{d}}_1,\bm{\mathrm{d}}_2)
    ={}& \frac{4}{3\mathcal{N}_{\rm dss}}
    \langle d_{\bm{\mathrm{r}}_0}
    S_{\bm{\mathrm{r}}_0+\bm{\mathrm{d}}_1}
    \cdot
    S_{\bm{\mathrm{r}}_0+\bm{\mathrm{d}}_2}\rangle \\
    & -\frac{4}{3\mathcal{N}_{\rm ss}}
    \langle S_{\bm{\mathrm{r}}_0+\bm{\mathrm{d}}_1}
    \cdot
    S_{\bm{\mathrm{r}}_0+\bm{\mathrm{d}}_2}\rangle,
\end{aligned}
\label{eq:three_point_func}
\end{equation}
where $d_{\bm{\mathrm{r}}} = n_{\bm{\mathrm{r}}\uparrow}n_{\bm{\mathrm{r}}\downarrow}$ and $S_{\bm{\mathrm{r}}}$ denote the double-occupancy and spin operators at site $\bm{\mathrm{r}}$, respectively.
The normalization factors are $\mathcal{N}_{\rm dss}=\langle d\rangle\langle p\rangle^2$ and $\mathcal{N}_{\rm ss}=\langle p\rangle^2$, where $\langle d\rangle$ and $\langle p\rangle$ denote the average probabilities of double and single occupancy, respectively. The spatial distribution of $\mathcal{C}_{\rm dss}$ is shown in Fig.~\ref{Fig:FM_TL_Hubbard}(a).

We begin by evaluating $\mathcal{C}_{\rm dss}$ for nearest-neighbor spin pairs satisfying $\left| \bm{\mathrm{d}}_2 - \bm{\mathrm{d}}_1\right| = 1$ on the finite triangular-lattice cluster with a hexagonal boundary shown in Fig.~\ref{Fig:FM_TL_Hubbard}(a). In this geometry, $\bm{\mathrm{r}}_0$ is chosen to be the center of the cluster. The stoMPS simulation is performed at doping $\delta = 0.16$ (here $\delta = n-1$, and $n = N_{\rm e}/N$ is the average particle density). Our results reveal a pronounced ferromagnetic correlation ($\mathcal{C}_{\rm dss} > 0$) within the innermost shell of neighbors surrounding the dopant, closely resembling the Nagaoka polaron structure reported in Ref.~\cite{lebrat_observation_2024}. Conversely, correlations between the innermost and second shells are antiferromagnetic ($\mathcal{C}_{\rm dss} < 0$), while the correlations within the second shell remain ferromagnetic but are significantly suppressed compared to the immediate vicinity of the dopant. This spatial modulation is qualitatively consistent with recent findings from quantum gas microscopy and DQMC simulations~\cite{lebrat_observation_2024}.

Next, we extend our analysis to the YC $4\times12$ triangular-lattice cylinder and show the results in Figs.~\ref{Fig:FM_TL_Hubbard}(b, c).
We compute the averaged nearest-neighbor spin-spin correlation
\begin{equation}
\mathcal{C}_{\rm ss}^{\rm tot}
    = \frac{4}{9N_{\rm bulk}\mathcal{N}_{\rm ss}}
    \sum_{\bm{\mathrm{r}}\in\mathrm{bulk}}\sum_{\nu=1}^{3}
    \langle S_{\bm{\mathrm{r}}}\cdot
    S_{\bm{\mathrm{r}}+\bm{\mathrm{e}}_\nu}\rangle,
\end{equation}
where $\bm{\mathrm{e}}_\nu$ ($\nu=1,2,3$) are three distinct bond orientations of the triangular lattice, as shown in Fig.~\ref{Fig:FM_TL_Hubbard}(c). To reduce boundary effects, we choose the central $L/3$ columns as the bulk region.

Fig.~\ref{Fig:FM_TL_Hubbard}(b) compares stoMPS results at $U=12$ and $T=0.4$ with experimental data at $U=9.9(3)$ and $T=0.62(4)$.
Despite these parameter differences, the calculated doping dependence is qualitatively consistent with experiment.
The horizontal axis uses the bulk doping $\delta$ evaluated at $T=0.4$.
Here, $\mathcal{C}_{\rm ss}^{\rm tot}$ changes sign in the electron-doped regime, indicating a crossover to positive nearest-neighbor correlations.
The label $\delta_{\rm FM}$ marks the doping at this sign change.

Finally, Fig.~\ref{Fig:FM_TL_Hubbard}(c) illustrates the temperature dependence of $\mathcal{C}_{\rm ss}^{\rm tot}$ across various doping levels. Each curve corresponds to a fixed total particle number, while the bulk doping can vary with temperature. Upon cooling, lower-doping curves develop stronger negative correlations, whereas more strongly electron-doped curves cross over to positive correlations. Our work demonstrates that stoMPS provides an accurate and efficient method to access the low-temperature physics of the doped triangular-lattice Hubbard model, serving as a valuable guide for ongoing quantum simulation experiments.

\section{Optical Conductivity of the Hubbard Model at Finite Temperature}
\label{Sec:conductivity}
Beyond static equilibrium properties, the stoMPS framework can be naturally extended to compute finite-temperature dynamical and transport properties via real-time evolution of the Born samples. Given an ensemble of Born-sampled MPSs $\{\ket{\psi_\alpha(\beta)}\}$ satisfying $\mathbb{E}[\ket{\psi_\alpha(\beta)}\bra{\psi_\alpha(\beta)}] \equiv  \sum_\alpha p(\alpha)\, [\ket{\psi_\alpha(\beta)}\bra{\psi_\alpha(\beta)}]  = \rho(\beta)$, the thermal average of a dynamical two-point correlation function $\expval{A(t)B}$ is given by
\begin{equation}
\begin{aligned}
\expval{A(t)B} = \frac{\mathbb{E}[\bra{\psi_\alpha(\beta)}e^{i Ht}Ae^{-i Ht}B\ket{\psi_\alpha(\beta)}]}{\mathbb{E}[\langle\psi_\alpha(\beta)|\psi_\alpha(\beta)\rangle]}.
\end{aligned}
\end{equation}
The states $\ket{\psi_\alpha(\beta)}$ can be either Born samples from the supervector, or regenerated by the annealed Born resampling (App.~\ref{App:resample}).
In practice, for each Born sample $\alpha$, we simultaneously evolve two MPSs, $\ket{u(t)} \equiv e^{-i Ht}\ket{\psi_\alpha(\beta)}$ and $\ket{v(t)} \equiv e^{-i Ht}B\,\ket{\psi_\alpha(\beta)}$, in real time using TDVP~\cite{TDVP2011, TDVP2016}. The correlation function is then evaluated from the overlap $\bra{u(t)}A\ket{v(t)}$ at each time step and accumulated over the sample ensemble.

\begin{figure}[tbp]
\centering
\includegraphics[width=\linewidth]{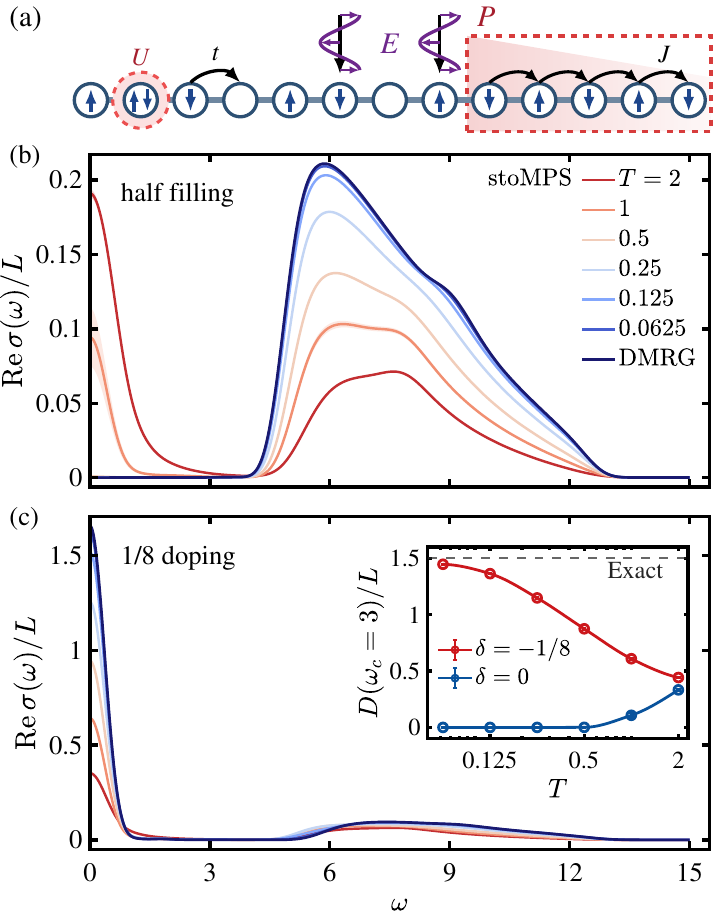}
\caption{Finite-temperature optical conductivity.
(a) Schematic of the setup for the 1D Hubbard chain ($t=1, U=8, L=64$). Real part of the optical conductivity $\mathrm{Re}\,\sigma(\omega)/L$ for (b) the half-filled ($N_{\rm e}=64$) and (c) $1/8$-hole-doped ($N_{\rm e}=56$) chain at representative temperatures. The inset of (c) shows the temperature evolution of the estimated Drude weight $D(\omega_c = 3)/L$ for $\delta = -1/8$ and $\delta = 0$, with the thermodynamic-limit Bethe-ansatz value at $T=0$ also indicated. All dynamical simulations are performed with bond dimension $D = 1024$ and maximum evolution time $t_{\rm max} = 10$. Finite-temperature results are averaged over $N_{\rm s} = 500$ Born samples taken at $T_0 = 2$. During cooling, the annealed resampling scheme described in App.~\ref{App:resample} is applied whenever $\eta<0.55$. Standard errors are represented by shaded bands or error bars.}
\label{Fig:conductivity}
\end{figure}

As a demonstration, we calculate the optical conductivity $\sigma(\omega)$ of the 1D Hubbard model, which constitutes a dynamical quantity measurable in ultracold-atom experiments~\cite{Anderson2019, Brown2019}.
As illustrated in the schematic of Fig.~\ref{Fig:conductivity}(a), an applied electric field $E$ couples to the system via $-E P$ in the scalar potential gauge, where $P = \sum_i x_i n_i$ is the polarization operator ($x_i$ being the site coordinate). According to linear-response theory, the optical conductivity is given by
\begin{equation}
    \sigma(\omega) = i\int_{0}^{\infty} \mathrm{d}t\, e^{i\omega t}\expval{[J(t), P]}_{\beta},
\end{equation}
where $J = i[H, P] = -it \sum_{i,\sigma} (c_{i\sigma}^\dagger c_{i+1,\sigma}^{\phantom{\dagger}} - \text{h.c.})$ is the total current operator.
We note that the current--polarization formulation adopted here is equivalent to the conventional current--current Kubo formula in the vector-potential gauge~\cite{Kubo1957}. To suppress the nonphysical oscillations arising from truncating the real-time evolution at $t_{\rm max} = 10$, we apply a compactly supported Parzen window function in the Fourier transform, which results in a frequency resolution of $\Delta \omega \approx 0.4$~\cite{Kuhner1999, Li2022, Chen2026}.

The present choice is convenient and advantageous under open boundary conditions, where the position operator, and hence the polarization $P$, is well defined. Under periodic boundary conditions, in contrast, $P$ becomes ill defined and the current--current formulation is usually invoked. Importantly, the present formulation yields the Drude weight directly as a coherent low-frequency peak, circumventing both the explicit $1/\omega$ prefactor and the delicate cancellation between diamagnetic and paramagnetic contributions encountered in the current--current Kubo formula \footnote{For the current--current formula, the Drude weight instead enters as a singular term, $\mathrm{Re}\,\sigma(\omega) = D\delta(\omega) + \sigma_{\rm reg}(\omega)$, with $D = \pi[\expval{-K}_\beta - \Lambda(0)]$ composed of the diamagnetic term $\expval{-K}_\beta$ ($K$ being the kinetic-energy operator) and the paramagnetic current response $\Lambda(\omega) = i\int_0^\infty \mathrm{d}t\, e^{i\omega t}\expval{[J(t),J(0)]}_\beta$.}.

Figures~\ref{Fig:conductivity}(b) and \ref{Fig:conductivity}(c) present the real part of the optical conductivity $\mathrm{Re}\,\sigma(\omega)/L$ for the half-filled and $1/8$-hole-doped Hubbard chain, respectively. For the half-filled chain [Fig.~\ref{Fig:conductivity}(b)], the high-temperature response ($T \gtrsim 1$) features a thermally activated Drude-like peak near zero frequency, reflecting the high-temperature metallicity. As temperature decreases, this low-frequency peak vanishes and spectral weight is transferred to higher frequencies ($4 \lesssim \omega \lesssim 13$) to form an incoherent absorption band of doublon--holon excitations across the Mott gap. At sufficiently low temperature ($T/t \simeq 0.0625$), our stoMPS results agree excellently with the ground-state DMRG data in Fig.~\ref{Fig:conductivity}(b), as the system is an exact Mott insulator with vanishing Drude weight ($D=0$)~\cite{Lieb1968PRLAbsence, Stafford1991PRBFiniteSize}. In contrast, introducing hole doping fundamentally alters the transport behavior [Fig.~\ref{Fig:conductivity}(c)]. The Drude peak grows monotonically upon cooling, indicating the emergence of coherent metallic quasiparticles at low temperatures.

To quantitatively trace the temperature evolution of the charge transport, we evaluate the effective Drude weight $D(\omega_c) = \int_{-\omega_c}^{\omega_c} \mathrm{Re}\,\sigma(\omega)\,\mathrm{d}\omega$ with an energy cutoff $\omega_c = 3$, shown in the inset of Fig.~\ref{Fig:conductivity}(c). For the half-filled insulator, $D(\omega_c)/L$ is strongly suppressed upon cooling and approaches zero as $T \to 0$, faithfully capturing the thermal freeze-out of charge carriers. For the doped metal, $D(\omega_c)/L$ increases monotonically with decreasing temperature and saturates to the exact Bethe ansatz result in the thermodynamic limit ($D_{\rm BA}/L \simeq 1.50$)~\cite{Kawakami1991PRBConductivity, Shastry1990PRLTwisted, Stafford1993PRBScaling}. These results demonstrate that stoMPS provides a powerful new approach for simulating many-body dynamics at finite temperatures.

\section{Stochastic PEPS for 2D Lattices}
\label{Sec:stoPEPS}
The Born sampling construction admits a direct 2D PEPS generalization. In this section, we propose the stoPEPS algorithm and apply it to a prototype quantum lattice model, i.e., quantum Ising model (QIM) described by the Hamiltonian
\begin{equation}
H_{\rm TFIM} = -J \sum_{\langle i,j \rangle} S^z_i S^z_j - B \sum_i S_i^x.
\label{Eq:TFIM}
\end{equation}
Here $S_i^{x,z}$ are two components of the spin-$1/2$ operators and $\langle i,j\rangle$ denotes nearest-neighbor bonds. We set the exchange coupling $J=1$ as the energy unit and consider the QIM at its quantum critical point (QCP) driven by transverse field $B_c\simeq 1.52$~\cite{Blote2002}. The simulations are performed on a finite $L_x\times L_y$ square lattice, with open boundary conditions on both directions.

\subsection{Born sampling of stoPEPS}
Following the purification construction of Eq.~\eqref{Eq:R}, the supervector for mixed state at inverse temperature $\beta$ reads
\begin{equation}
    \dket{\Psi(\beta)} = \sum_{\{s_i\}, \{\alpha_i\}} \mathrm{tTr}\!\left[ T^{[1]}_{s_1 \alpha_1} \cdots T^{[N]}_{s_N \alpha_N} \right] \ket{s_1, \dots, s_N}\otimes \ket{\alpha_1, \dots, \alpha_N}.
    \label{Eq:PEPO}
\end{equation}
$T^{[i]}_{s_i \alpha_i}$ denotes the local tensor at site $i$, carrying a physical index $s_i$, an auxiliary index $\alpha_i$, and virtual indices of bond dimension $D$ contracted according to the lattice geometry, and $\mathrm{tTr}$ denotes the trace over all contracted virtual bonds~\cite{Verstraete2004PEPS, Orus2014, Cirac2021RMP}.

\begin{figure}[t!]
\includegraphics[width=\linewidth]{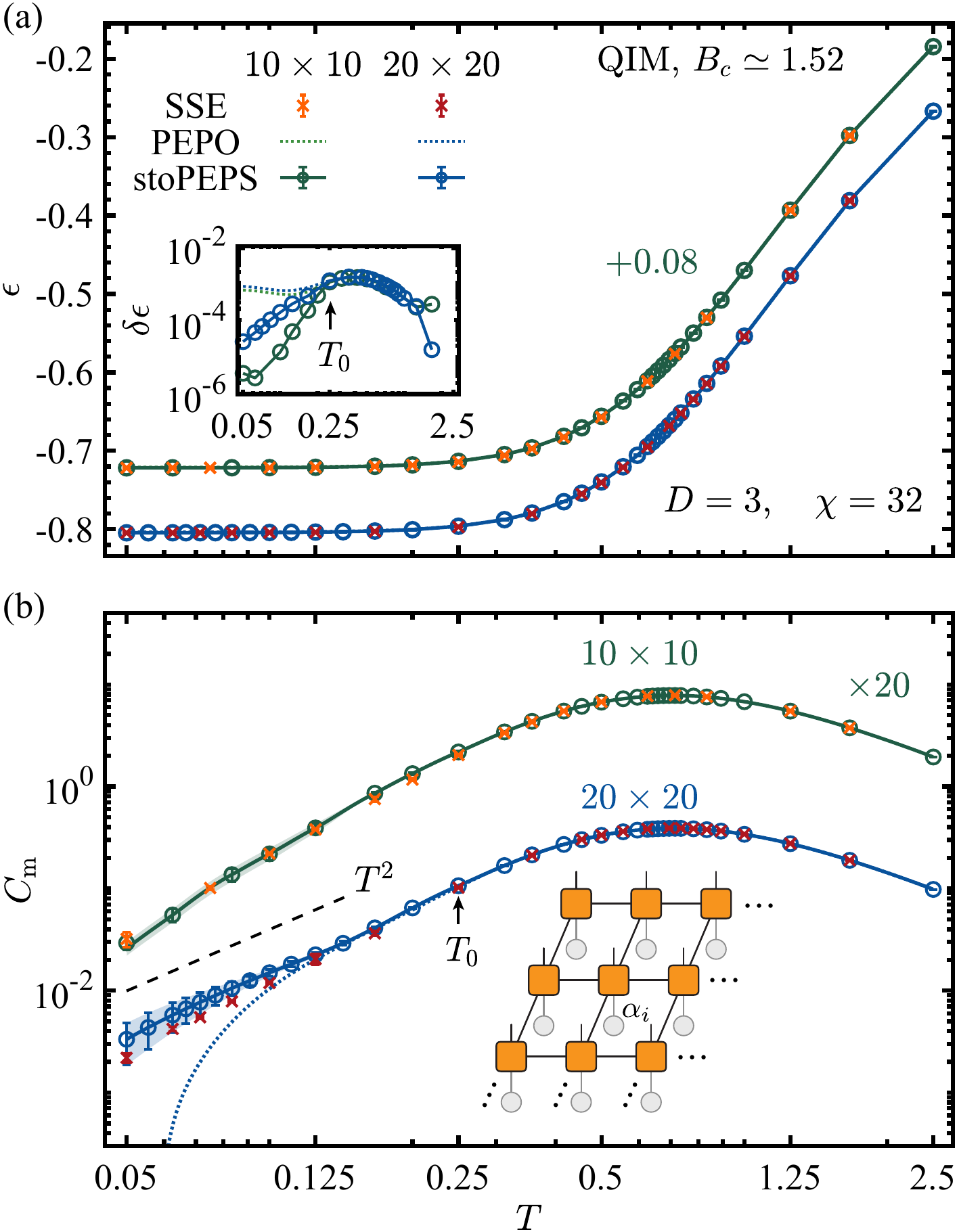}
\caption{stoPEPS results on the 2D QIM on $10\times10$ and $20\times 20$ square lattices with open boundaries.
(a) Energy per site $\epsilon$ at the QCP $B_c \simeq 1.52$, where the $10\times10$ data are shifted upward by 0.08 for clarity. The simulations are conducted with stoPEPS, PEPO purification, and SSE (see App.~\ref{App:fu-stoPEPS}). Both the PEPO and stoPEPS results use the bond dimension $D=3$, and the boundary-MPS dimension $\chi=32 \gg D^2$ well converges the results; the stoPEPS data are obtained with $N_{\rm s} = 200$ ($N_{\rm s}=1000$) Born samples initialized at $T_0=0.25$ for $10\times 10$ ($20\times 20$) lattice. Inset shows the energy deviation $\delta\epsilon$ of stoPEPS and PEPO from the SSE reference.
(b) Magnetic specific heat per site $C_{\rm m}$.  The dashed line indicates $C_{\rm m} \propto T^2$, and the arrow marks the Born sampling temperature $T_0$,
where the $10\times10$ data are multiplied by 20. The shaded area indicates the statistical error of stoPEPS. Inset shows a PEPS sample generated from the PEPO.}
\label{Fig6}
\end{figure}

In practice, we represent the supervector $\dket{\Psi(\beta_0)}$ in the equivalent form of a PEPO [see the inset of Fig.~\ref{Fig6}(b)], prepared at an intermediate inverse temperature $\beta_0$. We start from the identity and evolve the network with a sequence of two-site imaginary-time gates~\cite{Li2011, Czarnik2012PEPS} from the second-order Trotter decomposition.
All imaginary-time evolution is carried out with the full-update method~\cite{JordanEtAl2008, LubaschEtAl2014, PhienEtAl2015}. After a two-site gate enlarges a bond, the updated tensor pair is compressed in the environment of the entire finite 2D network, rather than in the local environment as in the simple update~\cite{Jiang2008PRL, Li2012Bethe}. The network environment is contracted with boundary MPS of bond dimension $\chi$ (see details in App.~\ref{App:fu-stoPEPS}).

Born sampling then proceeds as in Sec.~\ref{Sec:Alg}. We choose a complete orthonormal product basis $\ket{\alpha} = \bigotimes_i \ket{\alpha_i}$ of the auxiliary space; in the spin-$1/2$ QIM below, we use the eigenbasis of $S^x$, $\ket{\alpha_i} = (\ket{0}+\alpha_i\ket{1})/\sqrt{2}$ with $\alpha_i=\pm 1$.  Projecting each auxiliary index onto $\bra{\alpha_i}$ collapses the PEPO into a PEPS sample [see inset of Fig.~\ref{Fig6}(b)], according to Eq.~\eqref{Eq:Born}. Upon further cooling, each PEPS sample follows Eq.~\eqref{Eq:continue}, and thermal expectation values are estimated through the weighted ratio of Eq.~\eqref{Eq:ratio}. Here, the conditional sampling probabilities and the observable expectation values are also evaluated with the full update contraction.

\subsection{Results on the square-lattice quantum Ising model}
Figure~\ref{Fig6}(a) shows the energy per site $\epsilon$ of the $10\times 10$ and $20\times 20$ quantum Ising lattices versus temperature. Both the stoPEPS and PEPO results, at the same bond dimension $D=3$, agree with SSE quantum Monte Carlo data~\cite{Sandvik1991, Sandvik2010} over the entire temperature range. The SSE data are taken as a reference for comparing the accuracy of stoPEPS and PEPO results: the PEPO error is around $10^{-3}$, whereas the stoPEPS error is smaller by orders of magnitude upon cooling.

In Fig.~\ref{Fig6}(b), we show the magnetic specific heat $C_{\rm m}$ results, where the advantage of stoPEPS becomes more evident. Below $T_0 = 0.25$, the stoPEPS results agree with the SSE data and exhibit an approximate algebraic scaling. In contrast, the PEPO specific heat decays too rapidly and deviates from the SSE reference below $T \simeq 0.125$. This comparison highlights that stoPEPS generates accurate results at low temperatures, where the purification-based PEPO approach with the same $D$  already loses precision. These results demonstrate the capability of stoPEPS on 2D spin models. Its generalization to the Hubbard and $t$-$J$ models is readily accessible, which may shed new light on outstanding questions surrounding high-$T_c$ superconductivity and intertwined orders in the finite-temperature phase diagrams~\cite{tanTRG2023, Qu2022tJ, Qu2023bilayer}.

\section{Discussion}
\label{Sec:Discussion}
Viewed from a unified perspective, the present framework builds upon prevailing stochastic and purification approaches while resolving their inherent limitations. The METTS algorithm~\cite{White2009METTS, Stoudenmire2010} is the closest relative: each Born sample is precisely a METTS quantum state, and the stationary distribution of the METTS Markov chain coincides with the Born distribution. The essential distinction lies in the sampling mechanism: METTS generates samples through a Markov chain, which converges asymptotically at the price of autocorrelation, while Born sampling instead draws each state directly and independently from the supervector. The ergodicity in METTS Markov chain requires symmetry-breaking collapses, and symmetric variants restore the symmetry only via post-hoc projection of collapsed states, without accelerating the imaginary-time evolution itself~\cite{Bruognolo2015}. In Born sampling, the collapse basis are chosen as symmetry multiplets, thereby enjoying the full symmetry acceleration without Markov chains or autocorrelation. Therefore, the accessible system size can be pushed well beyond the state of the art: whereas METTS studies of the Hubbard model reached mostly width-4 cylinders~\cite{Wietek2021PRX, Sinha2025Forestalled} or $6\times6$ cluster~\cite{Sinha2024PEPSMETTS}, the present Born sampling attains width-10 cylinders with high accuracy.

The distinction to TPQ-MPS is equally clear. At $\beta_0=0$ the parent supervector reduces to the infinite-temperature identity, and the scheme reduces to a TPQ-MPS-like algorithm, which generates random states uniformly and without importance sampling~\cite{Iwaki2021, iwaki_sample_2024}: the thermal weights enter only as post-hoc reweighting factors, and the MPS restriction degrades typicality, leading to an algebraic growth of the required sample number with system size~\cite{iwaki_sample_2024}. The uniformly sampled random states are not symmetry eigenstates either, with symmetry acceleration absent. By contrast, sampling at $\beta_0>0$ concentrates the ensemble on thermally relevant configurations, substantially enhancing the sampling efficiency, with the annealed Born resampling maintaining the effective sample size upon cooling (\App{App:resample}). In this way, perfect Born sampling, realized as stoMPS and stoPEPS, achieves both symmetry acceleration and high sampling efficiency.

The present framework is also deeply rooted in the purification approach. The parent supervector $\dket{\Psi(\beta_0)}$ is prepared accurately and efficiently with state-of-the-art thermal tensor networks---tanTRG for the MPO realization and full-update PEPO for two dimensions~\cite{Li2011, Czarnik2012PEPS, Czarnik2015PEPS, Chen2018, tanTRG2023}---while below $T_0 = 1/\beta_0$, the Born samples, in MPS or PEPS form, take over down to ultralow temperatures. Our method thus covers precisely the temperature regimes where each scheme excels. Below $T_0$, stoMPS attains higher accuracy than purification at smaller bond dimensions and, thanks to the fewer physical indices, greatly reduces the evolution time per Born sample. Moreover, since the samples are independent, their evolution can be fully parallelized, reducing the computational wall-clock time (see App.~\ref{App:CBE-TDVP}).

\section{Summary and outlook}
\label{Sec:Sum}
In this work, we have developed the Born sampling framework for symmetric thermal tensor networks, realized as stoMPS and stoPEPS. The framework unites two long-sought but previously mutually exclusive advantages in a single scheme: importance sampling and symmetry acceleration. This is achieved by perfect Born sampling of the symmetry multiplets directly, performed at an intermediate temperature $\beta_0$. Evolving the low-entanglement MPS samples individually, instead of the full density-matrix MPO, yields higher accuracy with lower computational costs down to ultralow temperatures. With these features, we have performed high-precision finite-temperature simulations of the Hubbard cylinders up to the width $W=10$, revealed scalar chiral spin order and kinetic ferromagnetism in the triangular-lattice Hubbard model down to $T/t = 1/64$, computed the optical conductivity of Hubbard chains with the zero-frequency Drude peak clearly resolved, and studied quantum Ising lattices as large as $20\times20$. These advances demonstrate the power of the method.

Several intriguing directions are ready to be explored from here.
The Born samples can be evolved in real time or combined with Krylov-space techniques~\mbox{\cite{Kovalska2025TaSK}}, granting access to finite-temperature dynamics and transport properties. We demonstrate the optical conductivity in the present study, and leave more quantities, like the electronic and magnetic spectral functions, to future studies. Moreover, the generalization of stoPEPS to fermionic systems will extend the framework to fully scalable, large 2D lattices, providing a powerful new tool to map the finite-temperature phase diagram of the 2D Hubbard model.

Last but not least, the present framework holds particular significance for ultracold-atom quantum simulations of the Hubbard model.
First, with the canonical ensemble at fixed particle number, the calculations can be compared directly with experimental measurements, providing rigorous finite-temperature benchmarks and calibration~\mbox{\cite{Mazurenko2017, Hilker2017, Koepsell2021}}. Second, snapshots can be generated efficiently from the Born samples, providing direct access to snapshot-based observables in experiments, such as multi-point correlation functions. Third, it allows one to understand observations at the microscopic level and to propose new measurement protocols. Altogether, Born sampling grants access to arbitrary high-order correlations, rigorous finite-temperature benchmarks, and thermometry---capabilities increasingly demanded as experiments enter the cryogenic regime~\mbox{\cite{Xu2025Cryogenic}}. With accurate access to the pseudogap, strange-metal, and superconducting regimes~\mbox{\cite{Keimer2015Nature, Phillips2022, Qu2022tJ, Li2026PDW}}, the Born sampling method promises to advance the analysis of fermion Hubbard quantum simulations. In turn, it sheds new light on the long-standing open problems in the strongly correlated electrons and high-$T_c$ superconductivity.

\section*{Acknowledgement}
J.G., G.W., C.X., and W.L. acknowledge the \textit{\href{https://giggleliu.github.io/summer-school-2026/}{Harnessing Quantum 2026}} hackathon hosted by the Institute of Mathematics and Fundamental Physics (Hefei) for the stimulating and collaborative environment. W.L. thanks Lei Wang and Youjin Deng for insightful discussions. This work was supported by the National Natural Science Foundation of China (Grant Nos.~12625408, 12534009 and 12447101), the Innovation Program for Quantum Science and Technology (under Grant No. 2021ZD0301900), the China National Postdoctoral Program for Innovative Talents (Grant No. BX2026031). We thank the HPC cluster of ITP-CAS for technical support and the generous allocation of CPU time.
The development of this work made use of Codex (Sol and Astra, OpenAI) and Kimi Code (K3, Moonshot AI) for numerical techniques discussion, programming, figure preparation, and language polishing. The authors conceived the method and designed the workflow, performed the analysis, and take full responsibility for the scientific content of this work.

\section*{Data and Code availability}
The data that support the findings of this study are available from the corresponding author upon reasonable request.
The source code that supports the findings of this study is publicly available
in the GitHub repositories StoMPS.jl~\cite{StoMPS.jl} and StoPEPS.jl~\cite{StoPEPS.jl},
which are built upon and distributed as part of the open-source ThermoTN
project~\cite{ThermoTN}.

\appendix

\section{Perfect Born sampling algorithm}
\label{App:algorithm}

The workflow of the Born sampling algorithm is summarized in Alg.~\ref{alg:QMPS}.
The supervector $\dket{\Psi(\beta_0)}$ is first prepared in an MPO or PEPO representation at an intermediate inverse temperature $\beta_0$, e.g.,
using the linearized tensor renormalization group (LTRG)~\cite{Li2011, Dong2017},  exponential tensor renormalization group (XTRG)~\cite{Chen2018, Li2019}, tanTRG~\cite{tanTRG2023}, or PEPO-based imaginary-time evolution~\cite{Li2011, Czarnik2012PEPS, Czarnik2015PEPS, Czarnik2016TNR, Zhang2025Scalable}.
From the supervector $\dket{\Psi(\beta_0)}$,
$N_{\rm s}$ independent samples are generated.
Each sample, i.e., an auxiliary configuration $\alpha = (\alpha_1, \dots, \alpha_N)$, is drawn sequentially via the chain rule, with the conditional probability at each site evaluated from the local conditional reduced density matrix ${R}_{\beta_0}^{(i)}$ [Sec.~\ref{Sec:Alg} and Fig.~\ref{Fig1}(c)], yielding the normalized Born sample $\ket{\psi_{\alpha}(\beta_0)}$. Each sample is then evolved to the target inverse temperature $\beta$, $\ket{\tilde{\psi}_{\alpha}(\beta)} = e^{-(\beta-\beta_0)H/2}\ket{\psi_{\alpha}(\beta_0)}$, using the series-expansion thermal tensor network (SETTN)~\cite{Chen2017} or time-dependent variational principle (TDVP)~\cite{TDVP2011,TDVP2016}, yielding the accumulated weight $w_{\alpha} = \|\tilde{\psi}_{\alpha}(\beta)\|^2$ and the normalized state $\ket{\psi_{\alpha}(\beta)} = \ket{\tilde{\psi}_{\alpha}(\beta)}/\sqrt{w_\alpha}$. Observables are evaluated on the normalized states and accumulated with their weights, giving the thermal expectation value through Eq.~\eqref{Eq:ratio}.

To be more specific, we present below the construction of $R_\beta^{(i)}$ introduced in Sec.~\ref{Sec:Alg} and Fig.~\ref{Fig1}(c). Suppose the auxiliary indices of the first $(i-1)$ sites have already been drawn as $\alpha_{<i} = (\alpha_1, \dots, \alpha_{i-1})$. The local conditional reduced density matrix for the $i$-th site reads
\begin{equation}
    \begin{aligned}
        R_{\beta}^{(i)}(\alpha_{<i}) &\equiv
        \frac{1}{p(\alpha_{<i})}
        \mathrm{Tr}_{\rm phys} \mathrm{Tr}_{i+1, \dots, N}
\left[\langle \alpha_{<i} \dket{\Psi(\beta)}
\dbra{\Psi(\beta)} \alpha_{<i} \rangle\right] \\
&= \frac{1}{p(\alpha_{<i})}\mathrm{Tr}_{i+1, \dots, N} \left[ \bra{\alpha_{<i}} \rho(\beta) \ket{\alpha_{<i}} \right],
    \end{aligned}
\label{Eq:Rdef}
\end{equation}
where we have used $\mathrm{Tr}_{\rm phys}[\dket{\Psi(\beta)}\dbra{\Psi(\beta)}] = \rho(\beta)$, which follows from the symmetry between physical and auxiliary indices, and $\mathrm{Tr}_{i+1, \dots, N}$ denotes the partial trace over the auxiliary indices of sites $i+1, \dots, N$.
The auxiliary index $\alpha_i$ is then randomly drawn from the conditional probability
\begin{equation}
    \begin{aligned}
            p(\alpha_i|\alpha_{<i}) &\equiv \bra{\alpha_i} R_{\beta}^{(i)}(\alpha_{<i}) \ket{\alpha_i} \\
    &= \frac{1}{p(\alpha_{<i})} \mathrm{Tr}_{i+1, \dots, N}
\left[
    \langle \alpha_{<i}, \alpha_i |\rho(\beta)|\alpha_{<i}, \alpha_i \rangle
\right].
    \end{aligned}
\label{Eq:condp}
\end{equation}
Iterating this procedure from $i=1$ up to $N$, we obtain a configuration
$\alpha = (\alpha_1, \dots, \alpha_N)$ with the Born probability
\begin{equation}
    p(\alpha) = p(\alpha_1) p(\alpha_2|\alpha_1) \dots p(\alpha_N|\alpha_1, \dots, \alpha_{N-1}).
\label{Eq:chain}
\end{equation}

\begin{algorithm}[t]
\caption{Perfect Born Sampling}
\label{alg:QMPS}
Prepare the supervector $\dket{\Psi(\beta_0)}$ at $\beta_0$ in MPO or PEPO form\;
Initialize accumulators: $A_{\rm num} \gets 0$, $A_{\rm den} \gets 0$\;
\For{$k = 1$ \KwTo $N_{\rm s}$}{
    \tcp{Main Sampling Loop}
    \textbf{Step 1: Born Sampling at $\beta_0$}\;
    \For{$i = 1$ \KwTo $N$}{
    Evaluate the probability $p(\alpha_i|\alpha_{<i})$ from $R_{\beta_0}^{(i)}$ \tcp*{See Fig.~\ref{Fig1}(c) and App.~\ref{App:algorithm}}
    Sample $\alpha_i \sim p(\alpha_i | \alpha_{<i})$, normalize the tensor, and move to site $i+1$\;
    }
    Obtain the normalized Born sample $\ket{\psi_{k}(\beta_0)} \equiv \ket{\psi_\alpha(\beta_0)}$\;

    \vspace{1mm}
    \textbf{Step 2: Imaginary Time Evolution}\;
    $\ket{\tilde{\psi}_{k}(\beta)} \gets \exp[-(\beta-\beta_0) H / 2]\, \ket{\psi_{k}(\beta_0)}$ \;
    $w_{k} \gets \| \tilde{\psi}_{k}(\beta) \|^2$\;
    $\ket{\psi_{k}(\beta)} \gets \ket{\tilde{\psi}_{k}(\beta)} / \sqrt{w_{k}}$\;

    \vspace{1mm}
    \textbf{Step 3: Measurement}\;
    $O_{k} \gets \langle \psi_{k}(\beta)| O | \psi_{k}(\beta)\rangle$\;
    $A_{\rm num} \gets A_{\rm num} + O_{k} \cdot w_{k}$\;
    $A_{\rm den} \gets A_{\rm den} + w_{k}$\;
}
\Return $\langle O \rangle_\beta = A_{\rm num} / A_{\rm den}$\;
\end{algorithm}

\section{Annealed Born resampling}
\label{App:resample}

In this appendix, we present the annealed Born resampling scheme to effectively restore the sampling efficiency. For the sake of simplicity, we consider the case of Born sampling initialized at $\beta_0=0$; upon imaginary-time cooling to $\beta$, the sampling efficiency reads
\begin{equation}
\eta(\beta) \equiv \frac{
    \mathbb{E}_{\alpha \sim P_0(\alpha)}^2[w_\alpha(\beta)]}{
    \mathbb{E}_{\alpha \sim P_0(\alpha)}[w_\alpha^2(\beta)] }
    = \frac{\tilde{Z}(\beta)^2}{\mathbb{E}_{\alpha \sim P_0(\alpha)}[w_\alpha^2(\beta)]
},
\end{equation}
where $\tilde{Z}(\beta) = Z(\beta)/d^N$ with $d$ the local Hilbert space dimension.
In practice, $\mathbb{E}_{\alpha \sim P_0}$ is evaluated with the average over $N_{\rm s}$ samples, i.e., $\eta(\beta) \simeq \big[\sum_{k=1}^{N_{\rm s}} w_{k}(\beta)\big]^2 \big/ \big[N_{\rm s} \sum_{k=1}^{N_{\rm s}} w_{k}^2(\beta)\big]$.
Here the importance weights $w_\alpha(\beta) \equiv \|\tilde{\psi}_\alpha(\beta)\|^2$ become increasingly inhomogeneous: $\eta(0)$ starts from unity, since all weights $w_\alpha$ are identical, and decreases upon cooling. $P_0(\alpha)$ is the uniform distribution at $\beta_0=0$, which can be replaced by other proposal distributions without affecting the unbiasedness.

To this end, whenever $\eta$ falls below a prescribed threshold, we resample with an improved proposal $P_\beta(\alpha)$, chosen to minimize the second moment and restore the sampling efficiency
\begin{equation}
J[P_\beta(\alpha)] \equiv \mathbb{E}_{\alpha \sim P_\beta(\alpha)}\left[\frac{w^2_\alpha(\beta)} {P^2_\beta(\alpha)} \right].
\label{EqB2}
\end{equation}
Here the first moment of the resampled weights is fixed, $\mathbb{E}_{\alpha\sim P_\beta(\alpha)}[w_\alpha(\beta)/P_\beta(\alpha)]=\sum_\alpha w_\alpha(\beta)=Z(\beta)$, regardless of the proposal, so minimizing $J$ indeed amounts to minimizing the weight variance. By the Cauchy-Schwarz inequality, $\big[\sum_\alpha w_\alpha(\beta)\big]^2\leq J\sum_\alpha P_\beta(\alpha)$, i.e., $J\geq Z(\beta)^2$ as $\sum_\alpha P_\beta(\alpha)=1$, where the equality holds if and only if $P_\beta(\alpha)\propto w_\alpha(\beta)$, in which case the resampled weights become uniform and $\eta=1$.

The optimal proposal is thus the ideal Born distribution at the current temperature, $p_{\beta}(\alpha)\equiv\|\langle\alpha|\Psi(\beta)\rangle\!\rangle\|^2$. In practice, however, one can only access the empirical distribution
\begin{equation}
P_\beta(\alpha; N_{\rm s})
\equiv
\frac{
\sum_{k = 1}^{N_{\rm s}} w_{k}(\beta)
\left|\langle\alpha|\psi_{k}(\beta)\rangle\right|^2
}
{
\sum_{k = 1}^{N_{\rm s}} w_{k}(\beta)},
\label{Eq:resample_proposal}
\end{equation}
where $\ket{\psi_{k}(\beta)}$ denotes the $k$-th
sample with auxiliary configuration $\alpha(k)$ and weight $w_{k}(\beta)$.
Numerically, we find that resampling directly from $P_\beta(\alpha; N_{\rm s})$ overfits the finite sample set, and therefore introduce a mixture proposal $P_\beta^{\rm mix}(\alpha) \equiv \lambda P_\beta(\alpha; N_{\rm s}) + (1-\lambda) p_{\beta_0}(\alpha)$, where the mixing parameter $0\leq\lambda\leq1$ regularizes the empirical proposal~\cite{Hesterberg1995Defensive, Chopin2002Sequential}.

In practice, the mixing parameter
$\lambda$ is determined adaptively by minimizing the second moment of the resampled weights Eq.~\eqref{EqB2}, and the admixture of $p_{\beta_0}$ prevents the resampled ensemble from being trapped in a restricted subset of symmetry sectors. Moreover, rather than discarding all the parent states in the Born samples, the regenerated samples are merged with the parent ensemble, from which $N_{\rm s}$ states are then drawn via importance sampling (Alg.~\ref{alg:resample}); this strategy preserves high-weight configurations while effectively suppressing weight fluctuations.

\begin{algorithm}[t]
\caption{Annealed Born resampling at $\beta$}
\label{alg:resample}
\SetKwInput{KwIn}{Input}
\KwIn{Parent ensemble $\Gamma = \{(\ket{\psi_{k}(\beta)},w_{k}(\beta))\}_{k=1}^{N_{\rm s}}$, and $\dket{\Psi(\beta_0)}$}
Construct proposal $P_\beta(\alpha;N_{\rm s})$ using \Eq{Eq:resample_proposal}\;
Choose $0<\lambda<1$ by minimizing the second moment of the resampled weights Eq.~\eqref{EqB2}\;
$P_\beta^{\rm mix}\gets\lambda P_\beta+(1-\lambda)p_{\beta_0}$\;
\For{$k=1$ \KwTo $N_{\rm new}$}{
    \tcp{Main Resampling Loop}
    \textbf{Step 1: Resample Configurations at $\beta$}\;
    Sample $\alpha(k)\sim P_\beta^{\rm mix}$ sequentially\;

    \vspace{1mm}
    \textbf{Step 2: Collapse $\alpha(k)$ to $\dket{\Psi(\beta_0)}$}\;
    $\ket{\tilde\psi_{k, {\rm new}}(\beta_0)}\gets\langle\alpha(k)|\Psi(\beta_0)\rangle\!\rangle/\sqrt{P_\beta^{\rm mix}(\alpha(k))}$\;

    \vspace{1mm}
    \textbf{Step 3: Imaginary Time Evolution}\;
    $\ket{\tilde\psi_{k, {\rm new}}(\beta)}\gets e^{-(\beta-\beta_0)H/2}\ket{\tilde\psi_{k, {\rm new}}(\beta_0)}$\;
    $w_{k, {\rm new}} \gets \| \tilde{\psi}_{k, {\rm new}}(\beta) \|^2$\;
    $\ket{\psi_{k, {\rm new}}(\beta)} \gets \ket{\tilde{\psi}_{k, {\rm new}}(\beta)} / \sqrt{w_{k, {\rm new}}}$\;
}
Regenerated samples $\Gamma_{\rm new}=\{(\ket{\psi_{k, {\rm new}}(\beta)},w_{k, {\rm new}})\}_{k=1}^{N_{\rm new}}$

Merge $\Gamma$ and $\Gamma_{\rm new}$ into
$\Gamma_{\rm merge}=\{(\ket{\psi_{k, {\rm merge}}(\beta)},w_{k, {\rm merge}})\}_{k=1}^{N_{\rm s} + N_{\rm new}}$\;
Normalized weights $\omega_{k}\gets w_{k, {\rm merge}}/\sum_{k'=1}^{N_{\rm s} + N_{\rm new}}w_{k', {\rm merge}}$\;

Draw $N_{\rm s}$ samples $\{\ket{\psi_{k}(\beta)}\}_{k=1}^{N_{\rm s}}$ from $\Gamma_{\rm merge}$ with probability $\omega_{k}$\;
\Return $\{\ket{\psi_{k}(\beta)}\}_{k=1}^{N_{\rm s}}$\;
\end{algorithm}

\begin{figure}[t]
\includegraphics[width=1.0\linewidth]{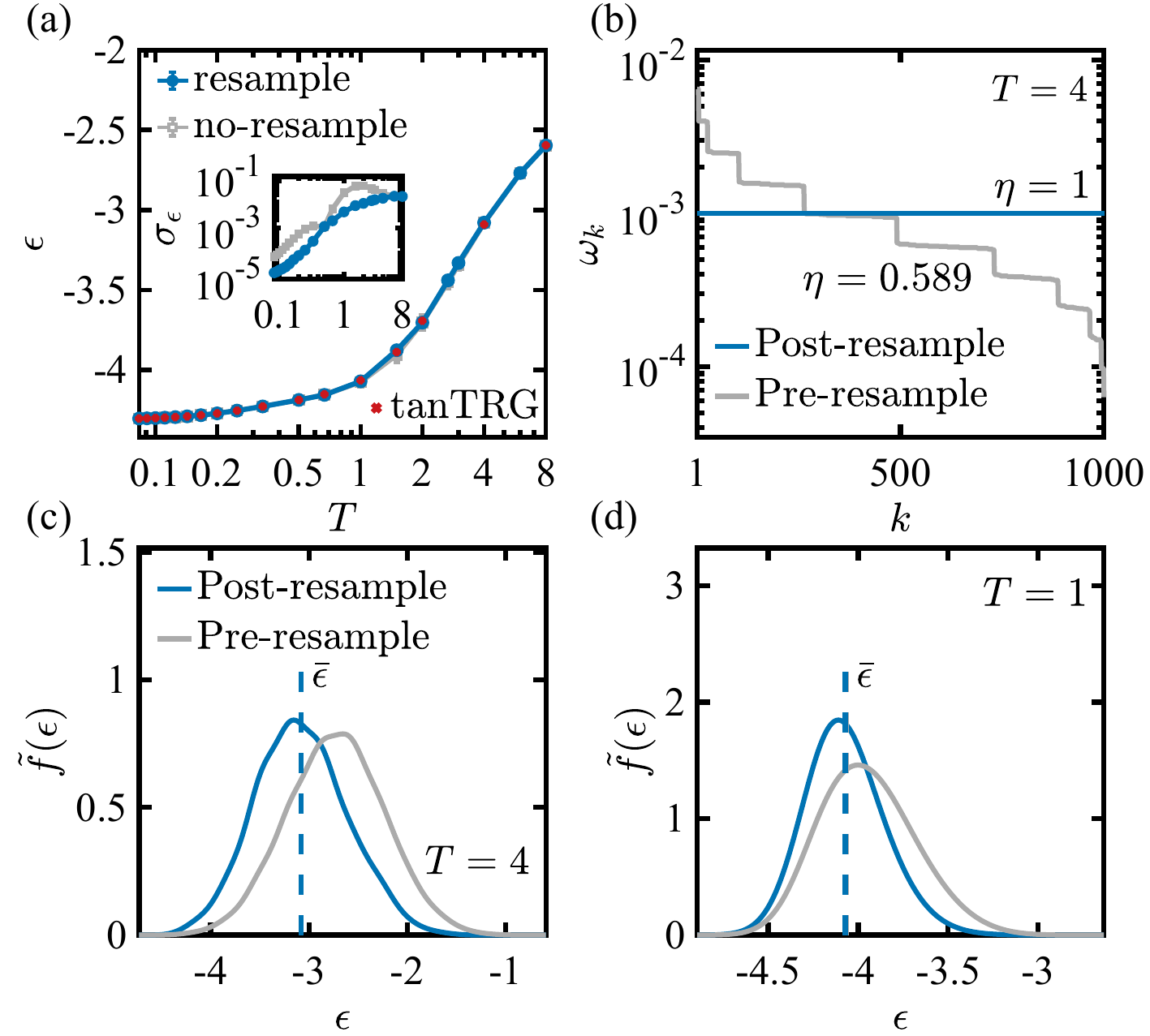}
\caption{Annealed Born resampling.
(a) Energy per site $\epsilon$ versus $T$ for the half-filled Hubbard chain ($L=12$) with open boundaries, $U=8$, and $\mathrm{SU(2)}_{\rm charge}\times\mathrm{SU(2)}_{\rm spin}$ symmetry. Blue circles and gray squares show stoMPS results with and without resampling, respectively, both using $N_{\rm s}=1000$ and perfect Born initialization at $T_0=8$, followed by cooling to $T=1/12$. Resampling uses $D=512$ and is performed at every temperature step during cooling; red crosses show tanTRG results with $D=1024$. The inset shows the standard error $\sigma_\epsilon$.
(b) Normalized sample weights $\omega_{k} = w_{k}/\sum_{k'=1}^{N_{\rm s}} w_{k'}$, sorted in descending order and indexed by $k$, before (gray) and after (blue) resampling at $T=4$. Resampling increases $\eta$ from $0.589$ to near unity.
(c, d) Unweighted kernel-density estimates $\tilde{f}(\epsilon)$ before (gray) and after (blue) resampling at $T=4$ and $1$, respectively. Vertical dashed lines indicate the weighted mean energy $\bar{\epsilon}$.
}
\label{Fig7}
\end{figure}

Figure~\ref{Fig7} demonstrates the performance of annealed Born resampling on the half-filled $L=12$ Hubbard chain, with the Born sampling initialized at $\beta_0=1/T_0=1/8$ in practice.
As shown in Fig.~\ref{Fig7}(a), the resampled stoMPS energies are in excellent agreement with the tanTRG benchmark, and their standard errors are smaller than those of the unresampled ensemble at the same sample size $N_{\rm s}=1000$ over the entire temperature range. Figure~\ref{Fig7}(b) illustrates the weight redistribution at $T=4$: resampling restores uniform weights and boosts the sampling efficiency $\eta \simeq 1$. Furthermore, the energy distributions $\tilde{f}(\epsilon)$ of samples in Figs.~\ref{Fig7}(c,d) show that resampling shifts the sample distribution towards the thermal expected value. These results demonstrate that resampling effectively reduces weight fluctuations and improves statistical precision in finite-temperature calculation.

\section{Evolution of Born samples and Benchmarks}
\label{App:CBE-TDVP}

After Born sampling at $\beta_0$ (or annealed Born resampling), each MPS sample is evolved to lower temperatures according to Eq.~\eqref{Eq:continue} using one-site TDVP with controlled bond expansion (CBE)~\cite{TDVP2011, TDVP2016, Gleis2023CBE, JhengWeionetdvp}.
Standard one-site TDVP evolves the state within the tangent space of an MPS manifold with fixed bond dimensions. CBE selectively enlarges the bond spaces to capture components of the evolution outside this tangent space, thereby reducing the projection error while retaining a computational cost close to that of one-site updates.

We compare the wall-clock times of stoMPS and tanTRG for the half-filled $U=8$ Hubbard model on the $4\times8$ cylinder [Fig.~\ref{Fig2}(a)], with both methods run on the same CPU cluster using 8 threads per calculation and imaginary-time TDVP, cooling from $T_0=1$ to $T=1/16$. At $T=1/4$, stoMPS with $D=12000$ achieves an energy accuracy comparable to that of tanTRG with $D=24000$. Over the cooling process, tanTRG requires $34.7~\mathrm{h}$, whereas the mean evolution time per stoMPS sample is $3.3~\mathrm{h}$---a tenfold speedup per task. Since the Born samples evolve independently and can be distributed across parallel tasks, switching from MPO evolution to MPS sample evolution significantly reduces the wall-clock time.

In the main text, we provide DQMC as numerically exact references at half filling. In practical calculations, we have up to $N_\mathrm{chain}=300$ Markov chains. Each chain includes $N_\mathrm{warm} = 1000$ warm up sweeps and $N_\mathrm{measure}=5000$ measurement sweeps. We then have $1500000$ measurements for each energy data shown in the main text. The imaginary time step is taken to be $\tau=0.05$, rendering the Trotter errors within $O(10^{-3})$.

\section{Born sampling with fixed particle number}
\label{App:fixed_particle}

In this appendix, we detail the deterministic conditional Born sampling scheme that strictly fixes the total particle number $N_{\rm e} = N_{\rm e}^{\rm target}$, realizing stoMPS simulations directly within the canonical ensemble.

Consider a purified supervector $\dket{\Psi(\beta_0)}$ of length $L$ at the initial temperature $\beta_0$, with the canonical center placed at the first site $i=1$. We perform sequential Born sampling from left to right along the chain. When sampling has proceeded up to site $i$, the auxiliary states on the preceding $(i-1)$ sites are fixed to $\alpha_{<i} \equiv (\alpha_1, \dots, \alpha_{i-1})$, carrying a cumulative particle number $N_{{\rm e},<i} \equiv \sum_{j=1}^{i-1} n(\alpha_j)$.

To strictly constrain the complete configuration $\alpha = (\alpha_1, \dots, \alpha_L)$ within the target charge sector $N_{\rm e} = N_{\rm e}^{\rm target}$, the auxiliary index $\alpha_i$ at site $i$ is drawn from the conditional probability
\begin{equation}
p(\alpha_i \mid \alpha_{<i}, N_{\rm e} = N_{\rm e}^{\rm target})
= \frac{p(\alpha_i, N_{\rm e} = N_{\rm e}^{\rm target} \mid \alpha_{<i})}{p(N_{\rm e} = N_{\rm e}^{\rm target} \mid \alpha_{<i})},
\label{Eq:cond_prob_ratio}
\end{equation}
where the denominator serves as the local normalization factor.
For each candidate state $\alpha_i$ carrying particle number $n(\alpha_i)$, the remaining $(L-i)$ sites must contain exactly $[N_{\rm e}^{\rm target} - N_{{\rm e},<i} - n(\alpha_i)]$ particles.
According to the Born rule, the unnormalized joint probability is evaluated by inserting the projector $P_{>i}(N_{\rm e}^{\rm target} - N_{{\rm e},\leq i})$ into the auxiliary indices of sites $i+1, \dots, L$, i.e.,
\begin{equation}
\begin{aligned}
& p(\alpha_i, N_{\rm e}\!=\!N_{\rm e}^{\rm target} \mid \alpha_{<i}) \\
& \quad \propto \vcenter{\hbox{
\begin{tikzpicture}[x=1pt, y=1pt, baseline={([yshift=-3pt]current bounding box.center)}]
  \def\ytop{20}
  \def\ybot{-20}
  \def\xone{0}
  \def\xi{46}
  \def\xnext{82}
  \def\xend{156}
  \fill[sitegreen, rounded corners=3pt] (\xi - 9, \ybot - 10) rectangle (\xi + 16, \ytop + 10);
  \draw[line width=0.6pt, color=black!65] (\xone, \ytop + 6) to[out=90, in=90, looseness=1.5] (\xone - 10, \ytop + 6)
       -- (\xone - 10, \ybot - 6) to[out=-90, in=-90, looseness=1.5] (\xone, \ybot - 6);
  \draw[line width=0.6pt, color=black!65] (\xi, \ytop + 6) to[out=90, in=90, looseness=1.5] (\xi - 9, \ytop + 6)
       -- (\xi - 9, \ybot - 6) to[out=-90, in=-90, looseness=1.5] (\xi, \ybot - 6);
  \draw[line width=0.6pt, color=black!65] (\xnext, \ytop + 6) to[out=90, in=90, looseness=1.5] (\xnext - 9, \ytop + 6)
       -- (\xnext - 9, \ybot - 6) to[out=-90, in=-90, looseness=1.5] (\xnext, \ybot - 6);
  \draw[line width=0.6pt, color=black!65] (\xend, \ytop + 6) to[out=90, in=90, looseness=1.5] (\xend + 10, \ytop + 6)
       -- (\xend + 10, \ybot - 6) to[out=-90, in=-90, looseness=1.5] (\xend, \ybot - 6);
  \draw[line width=0.8pt] (\xone + 6, \ytop) -- (\xone + 16, \ytop);
  \node[font=\small] at (\xone + 23, \ytop) {$\dots$};
  \draw[line width=0.8pt] (\xi - 16, \ytop) -- (\xi - 6, \ytop);
  \draw[line width=0.8pt] (\xi + 6, \ytop) -- (\xnext - 6, \ytop);
  \draw[line width=0.8pt] (\xnext + 6, \ytop) -- (\xnext + 22, \ytop);
  \node[font=\small] at (\xnext + 37, \ytop) {$\dots$};
  \draw[line width=0.8pt] (\xend - 22, \ytop) -- (\xend - 6, \ytop);
  \draw[line width=0.8pt] (\xone + 6, \ybot) -- (\xone + 16, \ybot);
  \node[font=\small] at (\xone + 23, \ybot) {$\dots$};
  \draw[line width=0.8pt] (\xi - 16, \ybot) -- (\xi - 6, \ybot);
  \draw[line width=0.8pt] (\xi + 6, \ybot) -- (\xnext - 6, \ybot);
  \draw[line width=0.8pt] (\xnext + 6, \ybot) -- (\xnext + 22, \ybot);
  \node[font=\small] at (\xnext + 37, \ybot) {$\dots$};
  \draw[line width=0.8pt] (\xend - 22, \ybot) -- (\xend - 6, \ybot);
  \foreach \x in {\xone, \xi, \xnext, \xend} {
    \draw[fill=mporange, rounded corners=2pt, line width=0.6pt] (\x - 6, \ytop - 6) rectangle (\x + 6, \ytop + 6);
    \draw[fill=mporange, rounded corners=2pt, line width=0.6pt] (\x - 6, \ybot - 6) rectangle (\x + 6, \ybot + 6);
  }
  \node[above, font=\small, text=black!85] at (\xone, \ytop + 10.5) {$1$};
  \node[above, font=\small, text=black!85] at (\xi, \ytop + 10.5) {$i$};
  \node[above, font=\small, text=black!85] at (\xnext, \ytop + 10.5) {$i\!+\!1$};
  \node[above, font=\small, text=black!85] at (\xend, \ytop + 10.5) {$L$};
  \draw[line width=0.8pt] (\xone, \ytop - 6) -- (\xone, 5.3);
  \draw[fill=projgray, draw=black, line width=0.5pt] (\xone, 3.3) circle (2.0pt);
  \draw[fill=projgray, draw=black, line width=0.5pt] (\xone, -3.3) circle (2.0pt);
  \draw[line width=0.8pt] (\xone, -5.3) -- (\xone, \ybot + 6);
  \node[right, font=\small] at (\xone + 2.5, 0) {$\alpha_1$};
  \node[font=\small] at (\xone + 23, 0) {$\dots$};
  \draw[line width=0.8pt] (\xi, \ytop - 6) -- (\xi, 5.3);
  \draw[fill=projgray, draw=black, line width=0.5pt] (\xi, 3.3) circle (2.0pt);
  \draw[fill=projgray, draw=black, line width=0.5pt] (\xi, -3.3) circle (2.0pt);
  \draw[line width=0.8pt] (\xi, -5.3) -- (\xi, \ybot + 6);
  \node[right, font=\small] at (\xi + 2.5, 0) {$\alpha_i$};
  \draw[fill=white, rounded corners=2.5pt, line width=0.7pt] (\xnext - 7, -6.8) rectangle (\xend + 7, 6.8);
  \node[font=\small] at ({(\xnext + \xend)/2}, 0) {$P_{>i}(N_{\rm e}^{\rm target} - N_{{\rm e},\leq i})$};
  \draw[line width=0.8pt] (\xnext, \ytop - 6) -- (\xnext, 6.8);
  \draw[line width=0.8pt] (\xnext, \ybot + 6) -- (\xnext, -6.8);
  \draw[line width=0.8pt] (\xend, \ytop - 6) -- (\xend, 6.8);
  \draw[line width=0.8pt] (\xend, \ybot + 6) -- (\xend, -6.8);
\end{tikzpicture}
}}
\end{aligned}
\label{Eq:fixed_N_diagram}
\end{equation}
As the projectors satisfy a recurrence relation
\begin{equation}
    P_{>i}(N_{\rm e}) = \sum_{\alpha_{i+1}} \ket{\alpha_{i+1}}\bra{\alpha_{i+1}}\otimes P_{>i+1}(N_{\rm e} - n(\alpha_{i+1})),
\end{equation}
the right-environments can be computed recursively via a right-to-left sweep within $O(D^3L^2)$ complexity. In practice, we first prepare the right-environments for all possible particle numbers and then reuse them for generating multiple independent samples.

An alternative approach to sampling within a fixed particle-number sector is post-selection (rejection sampling), where unconstrained grand-canonical Born samples are drawn and configurations with $\sum_i n(\alpha_i) \neq N_{\rm e}^{\rm target}$ are discarded. However, in an unconstrained thermal ensemble, the particle number distribution exhibits a Gaussian width $\sim \sqrt{L}$, so that the probability of hitting a specific $N_{\rm e}^{\rm target}$ scales as $\mathcal{O}(1/\sqrt{L})$ at best, and decays exponentially away from the mean filling. Rejection sampling thus suffers from catastrophic sample waste on large lattices. In contrast, the conditional Born sampling scheme deterministically guides every configuration into the target canonical sector with unity acceptance, incurring only a modest one-time overhead for precomputing the boundary projectors.

\begin{figure*}[t]
\includegraphics[width=1\linewidth]{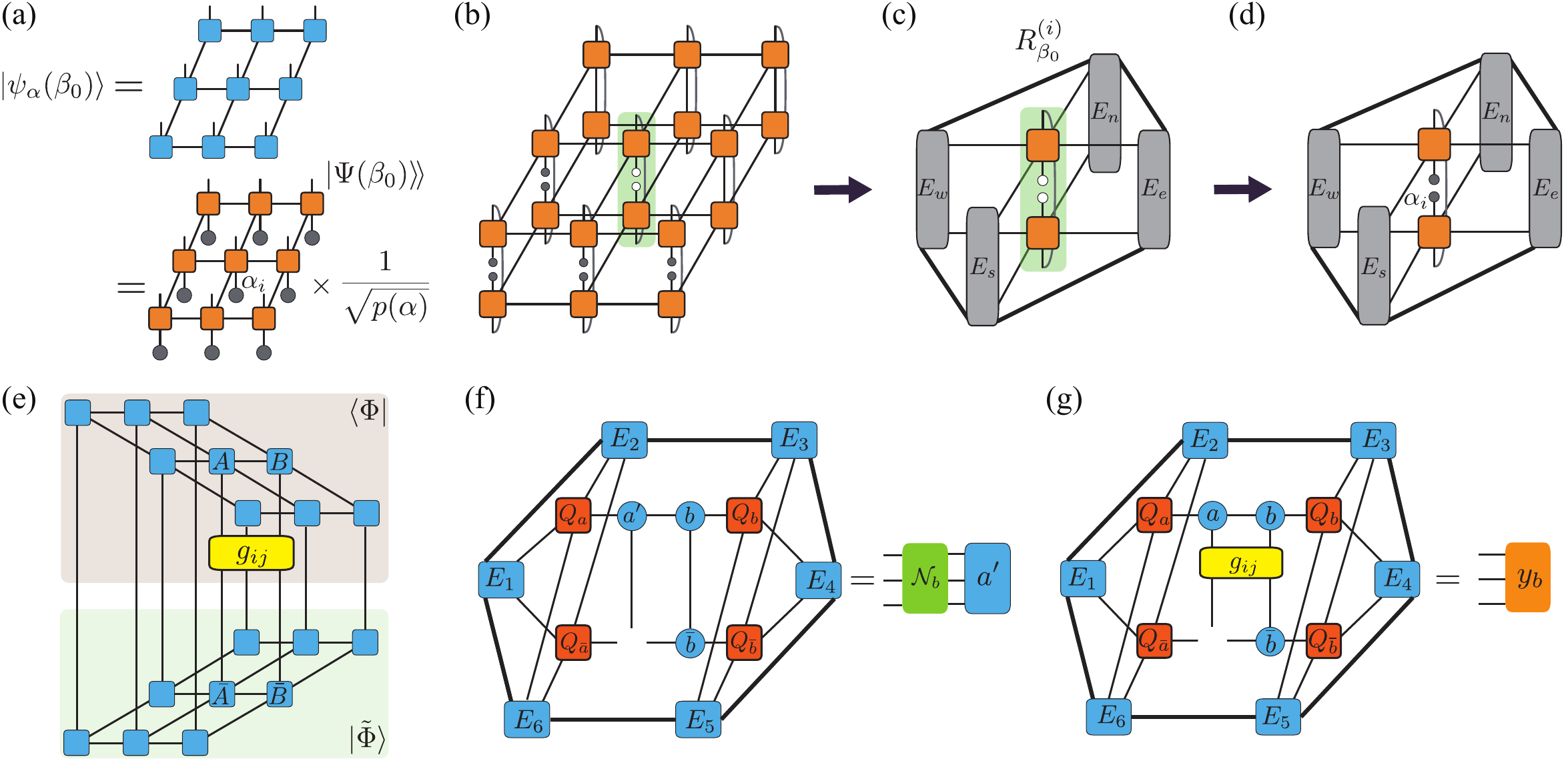}
\caption{
Overview of the stoPEPS algorithm.
(a) Generation of a Born sample $\ket{\psi_\alpha(\beta_0)}$ by projecting the auxiliary indices of the supervector $\dket{\Psi(\beta_0)}$ onto a sampled configuration $\alpha$.
(b--d) Sequential Born sampling at site $i$.
(b) double-layer network, where the filled circles represent the fixed preceding indices and the open circles represent the indices to be sampled.
(c) Local reduced density matrix $R_{\beta_0}^{(i)}$ obtained from the double-layer tensor contracted with boundary-MPS environments $E_w$, $E_n$, $E_e$, and $E_s$.
(d) The local auxiliary index is projected onto the sampled basis $\alpha_i$.
(e) Inner product $\langle\Phi|\tilde{\Phi}\rangle$ in the full update, where $|\tilde{\Phi}\rangle$ is the candidate global PEPS and $|\Phi\rangle=g_{ij}|\psi\rangle$ is the untruncated target.
(f) Contraction representing the action of the effective norm metric $\mathcal N_b$ on the candidate reduced core $a'$,
with the other core held fixed; the tensors $E_1,\ldots,E_6$ denote the environments.
(g) Contraction defining the corresponding right-hand side $y_b$ from the gate-applied target. Solving the normal equation $\mathcal N_b \bm a'=\bm y_b$ with $\bm a'=\operatorname{vec}(a')$ yields the updated core $a'$.
}
\label{Fig:Fig8}
\end{figure*}

\section{stoPEPS Algorithm for 2D Lattices}
\label{App:fu-stoPEPS}
In this appendix, we present the algorithmic details of the 2D stoPEPS scheme introduced in Sec.~\ref{Sec:stoPEPS}, whose construction follows the full-update PEPS formalism~\cite{JordanEtAl2008, LubaschEtAl2014, PhienEtAl2015}. In what follows, we denote $D$ as the retained virtual-bond dimension of the PEPO and PEPS, and $\chi$ for the boundary MPS used in contracting the 2D network.

We start from the identity operator, a trivial PEPO at infinite temperature. Acting successively with imaginary-time gates on its physical indices and compressing the enlarged virtual bonds yields a PEPO representation of the purified supervector $\dket{\Psi(\beta_0)}$ in Eq.~\eqref{Eq:PEPO}, on which the Born sampling is performed. The sampling scheme, imaginary-time evolution, and full-update compression are described in the following subsections.

\subsection{Born sampling of PEPO}
With the PEPO representation of the purified supervector $\dket{\Psi(\beta_0)}$ at hand, we now perform Born sampling on its auxiliary indices, as illustrated in Fig.~\ref{Fig:Fig8}(a). The auxiliary projection is carried out sequentially in a fixed row-by-row order. Suppose that the outcomes $\alpha_{<i}=(\alpha_1,\ldots,\alpha_{i-1})$ have already been selected. Contracting the remaining double-layer PEPO gives the reduced density matrix $R_{\beta_0}^{(i)}$ [Fig.~\ref{Fig:Fig8}(b)], from whose normalized diagonal elements $\alpha_i$ is drawn [Fig.~\ref{Fig:Fig8}(c)]; the auxiliary index is then projected onto the sampled state $\ket{\alpha_i}$ [Fig.~\ref{Fig:Fig8}(d)]. The sampling probability of the full configuration factorizes as $p(\alpha)=\prod_{i=1}^{N}p(\alpha_i\mid\alpha_{<i})$, with the auxiliary basis chosen as the $S^x$ eigenstates $\ket{\alpha_i}=(\ket{0}+\alpha_i\ket{1})/\sqrt{2}$, $\alpha_i=\pm1$.

As shown in Fig.~\ref{Fig:Fig8}(a), for each sampled configuration $\alpha$, projecting all auxiliary legs produces a stochastic PEPS, i.e., a Born sample, $\ket{\psi_\alpha(\beta_0)} = \langle\alpha|\Psi(\beta_0)\rangle\!\rangle/\sqrt{p(\alpha)}$. The ensemble average over the sampled configurations is then unbiased:
\begin{equation}
    \mathbb{E}_{\alpha\sim p}
    \left[\ket{\psi_\alpha(\beta_0)}\bra{\psi_\alpha(\beta_0)}\right]
    = \rho(\beta_0).
    \label{Eq:fu_stopeps_unbiased}
\end{equation}
Throughout this appendix, $p(\alpha)$ denotes the sampling probability, which is evaluated approximately via the boundary-MPS contraction of the 2D tensor network. As $\chi$ increases and the contraction converges, $p(\alpha)$ approaches the exact Born probability in Eq.~\eqref{Eq:Born}.

For a target inverse temperature $\beta\geq\beta_0$, with $\beta_0$ the inverse Born sampling temperature, each PEPS sample is further evolved in imaginary time as
\begin{equation}
    \ket{\tilde{\psi}_\alpha(\beta)}
    = e^{-(\beta-\beta_0)H/2}\ket{\psi_\alpha(\beta_0)}.
    \label{Eq:fu_stopeps_continue}
\end{equation}
Using Eq.~\eqref{Eq:fu_stopeps_unbiased}, the thermal expectation value can be evaluated directly from these evolved PEPS samples as
\begin{equation}
    \langle O\rangle_\beta
    = \frac{\mathbb{E}
    \left[\langle\tilde{\psi}_\alpha(\beta)|O
    \ket{\tilde{\psi}_\alpha(\beta)}\right]}
    {\mathbb{E}
    \left[\langle \tilde{\psi}_\alpha(\beta)|\tilde{\psi}_\alpha(\beta)\rangle\right]}.
    \label{Eq:fu_stopeps_observable}
\end{equation}
Both the numerator and denominator in Eq.~\eqref{Eq:fu_stopeps_observable} can be evaluated efficiently with boundary-MPS contraction (adopted in the present work)~\cite{JordanEtAl2008, Orus2014} or the corner transfer matrix approach~\cite{NishinoOkunishi1996, OrusVidal2009CTM}. The finite-$D$ truncation, finite-$\chi$ contraction, and Trotter errors are all systematically controlled.

\subsection{Imaginary-time evolution with full update}
On the square lattice with open boundary conditions, the transverse-field Ising Hamiltonian of Eq.~\eqref{Eq:TFIM} is split into nearest-neighbor terms,
\begin{equation}
    h_{ij} = -J S^z_i S^z_j - \frac{B}{z_i} S^x_i - \frac{B}{z_j} S^x_j,
    \label{Eq:fu_bond_ham}
\end{equation}
where $z_i$ is the coordination number of site $i$ and the field on each site is distributed evenly over its incident bonds, ensuring $\sum_{\langle i,j\rangle}h_{ij} = H$.
A two-site gate for a time increment $\delta\tau$ is
\begin{equation}
    g_{ij}(\delta\tau) = e^{-\delta\tau h_{ij}},
    \label{Eq:fu_gate}
\end{equation}
where $\delta \tau = 0.04. $
Each time step $\delta\tau$ is implemented using a second-order Trotter decomposition with a forward sweep followed by a reverse sweep. In the forward sweep, we update the horizontal bonds and then the vertical bonds, traversing each set from west to east and from north to south and applying $g_{ij}(\delta\tau/2)$ at each bond. We then apply the same half-step gates in the exact reverse bond order.

For a horizontal active bond, the left and right tensors are reduced exactly by thin QR and LQ factorizations,
\begin{equation}
    A = Q_a a,
    \qquad
    B = b Q_b,
    \label{Eq:fu_reduced}
\end{equation}
where $a$ and $b$ are the small cores varied during the local optimization, while the isometric tensors $Q_a$ and $Q_b$ are kept fixed. Contracting the gate with the active cores produces the target pair
\begin{equation}
    \Theta_{\alpha p_1 p_2 \beta} = \sum_{q_1 q_2 c} a_{\alpha q_1 c} b_{c q_2 \beta}\, g_{p_1 p_2, q_1 q_2},
    \label{Eq:fu_theta}
\end{equation}
where $g_{p_1p_2,q_1q_2}$ is the matrix representation of $g_{ij}$. A truncated SVD of $\Theta$ across $(\alpha,p_1)|(p_2,\beta)$ provides the initial candidate cores, which are subsequently improved by environment-weighted optimization.

The optimization minimizes the relative squared distance between the candidate global PEPS $\ket{\widetilde{\Phi}}$ and the untruncated target $\ket{\Phi} = g_{ij}\ket{\psi}$,
\begin{equation}
    \mathcal{C} = \frac{\bigl\| \ket{\widetilde{\Phi}} - \ket{\Phi} \bigr\|^2}{\langle \Phi | \Phi \rangle}.
    \label{Eq:fu_cost}
\end{equation}

Denoting the candidate reduced cores by $a'$ and $b'$, with $b'$ held fixed the cost function is quadratic in the vectorized core $\bm a' = \operatorname{vec}(a')$, leading to the stationary condition $\mathcal{N}_b \bm a' = \bm y_b$.
For numerical stability, we solve the regularized system
$\left(\mathcal{N}_b+\mu I\right)\bm a'=\bm y_b$,
where $\mathcal{N}_b$ is the environment-induced norm metric on the reduced parameters, $\bm y_b$ is the corresponding projection of the untruncated target, and the positive shift $\mu$ suppresses updates along null or poorly resolved gauge directions.

To obtain $\mathcal{N}_b$ and $\bm y_b$, the double-layer environments are effectively contracted by the boundary-MPS method~\cite{JordanEtAl2008, Orus2014} with truncated bond dimension $\chi$.
The truncation error of boundary-MPS contraction with $\chi = 32$ remains below $3 \times 10^{-8}$
($3 \times 10^{-5}$) for the $10\times10$ ($20\times20$) lattice
across all stochastic samples and
temperatures considered in our stoPEPS calculations.

The normal equation is solved by diagonally preconditioned conjugate gradients without assembling $\mathcal{N}_b$ explicitly; the left and right cores are then updated alternately until convergence.

\subsection{SSE benchmark}
For the reference benchmarks in Fig.~\ref{Fig6}, we perform SSE quantum Monte Carlo simulations~\cite{Sandvik1991, Sandvik2010} on the
$10\times10$ and $20\times20$ square-lattice quantum Ising model at the quantum critical point $B_c\simeq1.52$ with open boundary conditions. Each Monte Carlo sweep combines standard diagonal updates with cluster updates of the off-diagonal operator string.

For the largest $20 \times 20$ lattice, SSE simulations are carried out across 128 independent Markov chains. For each chain, $(0.5\text{--}2)\times 10^5$ warm-up sweeps are discarded for thermalization. The operator-string cutoff is dynamically adjusted to accommodate the largest sampled expansion order, eliminating truncation bias across all temperatures down to $T=0.05$. Production runs accumulated extensive statistics, with $5\times 10^7$ sweeps per temperature point for $T \geq 0.125$ and $3.2\times 10^{10}$ sweeps for $T < 0.125$. The specific heat per site is evaluated via the energy-fluctuation estimator $C_{\rm m} = \beta^2 (\langle H^2 \rangle - \langle H \rangle^2)/N$. Integrated autocorrelation times $\tau_{\rm int}$ for the energy are monitored using blocking analysis and remain below $\sim 6$ sweeps across all temperatures. Measurements are grouped into bins significantly larger than $\tau_{\rm int}$, and statistical errors are determined using jackknife resampling combined with inter-chain variance, reporting the larger of the two error estimates.

\bibliography{stoMPSRef}
\end{document}